\documentclass[12pt,epsf]{article}
\usepackage[pdftex]{graphicx}
\usepackage{amsmath,amssymb}
\usepackage{bm}
\usepackage{ascmac}
\usepackage{slashbox}
\graphicspath{{./fig/}}
\usepackage{cite}

\newcommand{\beq}{\begin{equation}}
\newcommand{\eeq}{\end{equation}}
\newcommand{\bea}{\begin{eqnarray}}
\newcommand{\eea}{\end{eqnarray}}
\newcommand{\non}{\nonumber\\}
\newcommand{\no}{\nonumber}
\newcommand{\ba}{\begin{array}}
\newcommand{\ea}{\end{array}}
\newcommand{\Slash}[1]{{\ooalign{\hfil/\hfil\crcr$#1$}}}

\def\lag{{\cal{L}}}

\def\pe2{p_E^2}

\begin{document}
\setlength{\baselineskip}{0.7cm}
\begin{titlepage} 
\begin{flushright}
OCU-PHYS 499  \\
NITEP 10
\end{flushright}
\vspace*{10mm}
\begin{center}{\LARGE\bf Fermion Mass Hierarchy \\
\vspace*{2mm}
in Grand Gauge-Higgs Unification}
\end{center}
\vspace*{10mm}
\begin{center}
{\large Nobuhito Maru}$^{a,b}$ and 
{\large Yoshiki Yatagai}$^{a}$, 
\end{center}
\vspace*{0.2cm}
\begin{center}
${}^{a}${\it 
Department of Mathematics and Physics, Osaka City University, \\ 
Osaka 558-8585, Japan}
\\
${}^{b}${\it Nambu Yoichiro Institute of Theoretical and Experimental Physics (NITEP), \\
Osaka City University, 
Osaka 558-8585, Japan} 
\end{center}
\vspace*{1cm}

\begin{abstract} 
Grand gauge-Higgs unification of five dimensional $SU(6)$ gauge theory 
 on an orbifold $S^1/Z_2$ is discussed. 
The Standard model (SM) fermions are introduced on one of the boundaries 
 and some massive bulk fields are also introduced 
 so that they couple to the SM fermions through the mass terms on the boundary. 
Integrating out the bulk fields generates the SM fermion masses 
 with exponentially small bulk mass dependences. 
The SM fermion masses except for top quark are shown to be reproduced 
 by mild tuning the bulk masses. 
One-loop Higgs potential is calculated 
 and it is shown that the electroweak symmetry breaking occurs 
 by introducing additional bulk fields.  
Higgs boson mass is also computed. 
\end{abstract}
\end{titlepage}

\section{Introduction} 
Gauge-Higgs unification (GHU) \cite{GH} is one of the attractive scenarios 
 among the physics beyond the Standard Model (SM), 
 which solves the hierarchy problem by identifying the SM Higgs field 
 with one of the extra spatial component of the higher dimensional gauge field. 
In this scenario, the most appealing feature is that physical observables in Higgs sector are calculable 
 and predictable regardless of the non-renormalizable theory. 
For instance, the radiative corrections to Higgs mass and Higgs potential are known to be finite 
 at one-loop \cite{1loop} and two-loop \cite{2loop} thanks to the higher dimensional gauge symmetry. 
Rich structures of the theory and its phenomenology have been investigated 
\cite{Higgsphys, GHUST, diphoton, Maru, GHUflavor, GHUmixing, triple, Yukawa, GHDM}.  

The hierarchy problem was originally addressed in grand unified theory (GUT) 
 as a problem how the discrepancy between the GUT scale and the weak scale are kept. 
Therefore, the extension of GHU to grand unification is an interesting direction to explore. 
The scenario of grand gauge-Higgs unification was discussed by one of the present authors 
 \cite{LM},\footnote{For earlier attempts and related recent works, see \cite{otherGGHU}} 
 where the five dimensional $SU(6)$ grand gauge-Higgs unification was considered 
 and the Standard Model (SM) fermions were embedded 
 in zero modes of some $SU(6)$ multiplets in the bulk. 
This embedding was very elegant in that it was a minimal matter content 
 without massless exotic fermions which is not included in the SM. 
That immediately means a minimal anomaly-free matter content. 
However, a crucial drawback was found that the down-type Yukawa couplings 
and the charged lepton Yukawa couplings are not allowed. 
This is because the left-handed quark (lepton) $SU(2)_L$ doublets 
 and the right-handed down quark (charged lepton) $SU(2)_L$ singlets 
 are embedded into different $SU(6)$ multiplets. 
As a result, Yukawa coupling in GHU originated from the gauge coupling cannot be allowed.  
This feature seems to be generic in GHU, therefore we have to give up 
 embedding all the SM fermions into the $SU(6)$ multiplets in the bulk 
 to obtain the SM Yukawa couplings. 
Fortunately, we know another approach to generate Yukawa coupling in a context of GHU \cite{CGM, SSS}.
In this approach, the SM fermions are introduced on the boundaries 
 (i.e. fixed point in an orbifold compactification). 
We also introduce massive bulk fermions, which couple to the SM fermions 
 through the mass terms on the boundary. 
Integrating out these massive fermions generates non-local SM fermion masses, 
 which are proportional to the bulk to boundary couplings and 
 exponentially sensitive to their bulk masses.  
Then, the SM fermion mass hierarchy can be obtained by very mild tuning of bulk masses. 

In this paper, we propose an improved $SU(6)$ grand GHU model \cite{LM}, 
 where the SM fermion mass hierarchy is obtained 
 by following the approach mentioned in the last paragraph. 
The SM fermions are introduced on the boundary as $SU(5)$ multiplets, 
 the four types of massive bulk fermions in $SU(6)$ multiplets 
 coupling to the SM fermions are introduced. 
We obtain the quark and lepton masses except for top quark 
 by integrating out the massive bulk fermions and tuning of the bulk masses. 
We also calculate one-loop Higgs potential 
 and study whether the electroweak symmetry breaking happens 
 and Higgs mass can be obtained. 
This issue is very nontrivial in GHU since the potential is generated at one-loop 
 and strongly depends on matter fermion content. 
We find that it is not possible to break the electroweak symmetry 
 by only the four types of bulk fermions. 
Then, we show that the electroweak symmetry breaking and a viable Higgs mass 
 can be realized by introducing additional bulk fermions with large dimensional representation. 

This paper is organized as follows. 
In the next section, we describe our model in detail. 
In section 3, the mechanism of the SM fermion mass generation is explained. 
It is shown that the SM fermion masses except top quark can be reproduced 
 by mild tuning of bulk masses. 
One-loop Higgs potential is calculated in section 3, 
 where the electroweak symmetry breaking and Higgs mass are analyzed. 
Section 4 is devoted to our conclusions and discussions. 
The details of calculations are summarized in Appendices. 
The branching rules of the representations relevant to our model are shown in Appendix A. 
In Appendix B, calculations of the Kaluza-Klein (KK) mass spectrum of bulk fields 
 are explained in some detail.

\section{Gauge and Higgs sector of our model} 
In this section, we briefly explain an $SU(6)$ GHU model \cite{LM}. 
We consider a five dimensional (5D) $SU(6)$ gauge theory 
 with an extra space compactified on an orbifold $S^1/Z_2$, 
 whose radius and coordinate are denoted by $R$ and $y$, respectively. 
$Z_2$ parities at each fixed points are given as follows. 
 \bea 
 P &=& \mbox{diag}(+,+,+,+,+,-) \, \, \mbox{at}~y=0, \non
 P' &=& \mbox{diag}(+,+,-,-,-,-) \, \, \mbox{at}~y=\pi R.
 \eea
We assign the $Z_2$ parity for the gauge field and the scalar field 
 as $A_{\mu}(-y)=PA_{\mu}(y)P^\dag$, $A_y(-y)=-PA_y(y)P^\dag$. 
Then, their fields have the following parities in components,  
 \beq
 A_{\mu} = \left(
    \ba{cc|ccc|c}
      (+,+) &(+,+) & (+,-) & (+,-) & (+,-) & (-,-) \\
      (+,+) &(+,+) & (+,-) & (+,-) & (+,-) & (-,-) \\
      \hline
      (+,-) &(+,-) & (+,+) & (+,+) & (+,+) & (-,+) \\
      (+,-) &(+,-) & (+,+) & (+,+) & (+,+) & (-,+) \\
      (+,-) &(+,-) & (+,+) & (+,+) & (+,+) & (-,+) \\
      \hline
      (-,-) &(-,-) & (-,+) & (-,+) & (-,+) & (+,+) 
    \ea
  \right),
 \eeq
 \beq
 A_{y} = \left(
    \ba{cc|ccc|c}
      (-,-) &(-,-) & (-,+) & (-,+) & (-,+) & (+,+) \\
      (-,-) &(-,-) & (-,+) & (-,+) & (-,+) & (+,+) \\
      \hline
      (-,+) &(-,+) & (-,-) & (-,-) & (-,-) & (+,-) \\
      (-,+) &(-,+) & (-,-) & (-,-) & (-,-) & (+,-) \\
      (-,+) &(-,+) & (-,-) & (-,-) & (-,-) & (+,-) \\
      \hline
      (+,+) &(+,+) & (+,-) & (+,-) & (+,-) & (-,-) 
    \ea
  \right),
 \eeq
where $(+,-)$ means that $Z_2$ parity is even (odd) at $y=0~(y = \pi R)$ boundary, for instance. 
We note that only the field with $(+, +)$ parity has a 4D massless zero mode ($n=0$) 
 as can be seen from the KK expansion in terms of mode function described in Appendix B. 
The $Z_2$ parity for $A_\mu$ indicates that $SU(6)$ gauge symmetry is broken 
 to $SU(3)_C \times SU(2)_L \times U(1)_Y \times U(1)_X$ 
 by the combination of the symmetry breaking pattern at each boundary,
 \bea 
 &&SU(6)\rightarrow SU(5)\times U(1)_X \, \, \mbox{at}~y=0, \\
 &&SU(6)\rightarrow SU(2)\times SU(4) \, \, \mbox{at}~y=\pi R.  
 \eea
The hypercharge $U(1)_Y$ is contained in Georgi-Glashow $SU(5)$ GUT, 
 which is an upper-left $5 \times 5$ submatrix of $6 \times 6$ matrix. 
Thus, we have
\bea
g_3 = g_2 = \sqrt{\frac{5}{3}}g_Y
\eea 
at the unification scale, which will not be so far from the compactification scale. 
$g_{3,2,Y}$ are the gauge coupling constants for $SU(3)_C, SU(2)_L, U(1)_Y$, respectively. 
This coupling relation implies that the weak mixing angle is the same as 
 that of Georgi-Glashow $SU(5)$ GUT model, $\sin^2 \theta_W=3/8$~($\theta_W:$weak mixing angle). 
This result can be explicitly checked for the bulk fermion in ${\bf 15}$ representation of $SU(6)$ 
 (see the next section).  
\bea
\sin^2 \theta_W = \frac{{\rm Tr}~I_3^2}{{\rm Tr}~Q^2} 
= \frac{((\frac{1}{2})^2 + (-\frac{1}{2})^2) \times 4}{((\frac{2}{3})^2 + (-\frac{2}{3})^2 + (\frac{1}{3})^2 
+ (\frac{1}{3})^2) \times 3 + 1^2 + 1^2} 
= \frac{3}{8}, 
\eea
where $I_3$ is the third component of $SU(2)_L$ isospin and $Q$ is an electric charge. 

$SU(2)_L$ Higgs doublet field is identified 
 with a part of an extra component of gauge field $A_y$.
 \beq 
 A_y=\frac{1}{\sqrt{2}}
  \left(\ba{c|c|c}
  \hspace{30pt}&\hspace{50pt}&H\\ \hline
  &&\\ 
  &&\\ \hline
  H^{\dag}&&\\  
  \ea\right). 
 \eeq
We suppose that a vacuum expectation value (VEV) of the Higgs field 
 is taken to be in the 28-th generator of $SU(6)$,  
 $\langle A_y^a \rangle = \frac{2\alpha}{Rg}\delta^{a\,28}$. 
$g$ is a 5D $SU(6)$ gauge coupling and $\alpha$ is a dimensionless constant. 
The VEV of Higgs field is given by $\langle H \rangle = \frac{\sqrt{2}\alpha}{Rg}$.
We note that the doublet-triplet splitting problem is solved by the orbifolding 
 since the $Z_2$ parity of the colored Higgs field is $(+, -)$ 
 and it become massive \cite{Kawamura}. 

Here we give some comments on $U(1)_X$ gauge symmetry 
 which remains unbroken by orbifolding. 
We first note that the $U(1)_X$ is anomalous as it stands 
 since the massless fermions are only the SM fermions 
 and their $U(1)_X $ charge assignments are not anomaly-free 
 (see Table \ref{table:representation} in the next section.).
However, it is easy to cancel the anomaly 
 by adding appropriate number of the SM singlet fermions with $U(1)_X$ charge only.   
In order to break the $U(1)_X$ spontaneously, 
 $U(1)_X$ charged scalars are introduced on the $y=0$ boundary for instance, 
 and we write down the potential of quadratic and quartic terms like the SM Higgs potential. 
Then, $U(1)_X$ is spontaneously broken by having the VEV for the scalars. 
  
\section{Fermion masses}
As mentioned in the introduction, 
 we have to give up embedding all the SM fermions into the $SU(6)$ multiplets in the bulk 
 to generate the fermion masses. 
The SM quarks and leptons are embedded into $SU(5)$ multiplets localized at $y=0$ boundary, 
 three sets of $\Psi_{10}$, $\Psi_{5^{\ast}}$, $\Psi_1$ along the sprit of GUT as much as possible. 
We also introduce various pair of bulk fermions $\Psi$ and $\tilde{\Psi}$ 
 with opposite $Z_2$ parities each other and constant mass term like $M\bar{\Psi} \tilde{\Psi}$ 
 in the bulk to avoid exotic massless fermions from them.  
$\tilde{\Psi}$ is referred as ``mirror fermions" in this paper. 
In this setup, we have no massless chiral fermions from the bulk and its mirror fermions. 
The massless fermions are the SM fermions only 
 and the gauge anomalies for the SM gauge groups are trivially canceled.  
In order to realize the SM fermion masses, 
 the boundary localized mass terms between the SM fermions localized at $y=0$ and the bulk fermions are necessary.  
To this end, we have to choose appropriate representations of $SU(6)$ for bulk fermions 
 so that the left(right)-handed fermion components in the bulk fermions couple to the right(left)-handed SM fermions 
 after the decomposition into the SM model gauge group representations. 
Note that the mirror fermions have no coupling to the SM fermions. 
Table \ref{table:representation} shows various representations for bulk and mirror fermions in our model 
 in addition to the SM fermions, which corresponds to the matter content for one generation. 
Totally, three copies of them are present in our model. 
 
 \begin{table}[h]
  \centering 
   \begin{tabular}{|c|c|c|} \hline
   bulk fermion
      & mirror fermion
         & SM fermion coupling to bulk \\ \hline
   $20^{*(-,-)} \supset Q_{20}^* (3^*,2)_{-1/6,3}^{(-,-)}, U_{20}^* (3^*,1)_{-2/3,-3}^{(+,+)}$
      &$20^{*(+,+)}$ 
         &$q^{\ast}_L(3^{\ast},2)_{-1/6,3},u^{\ast}_R(3^{\ast},1)_{-2/3,-3}$\\ \hline
   $56^{(-,+)} \supset Q_{56}(3,2)_{1/6,-3}^{(+,+)}, D_{56} (3,1)_{-1/3,-9}^{(-,-)}$   
      &$56^{(+,-)}$
         &$q_L(3,2)_{1/6,-3},d_R(3,1)_{-1/3,-9}$\\ \hline
   $15^{(+,+)} \supset L^*_{15} (1,2)_{1/2,-4}^{(-,-)}, E^*_{15} (1,1)_{1,2}^{(+,+)}$
      &$15^{(-,-)}$
         &$l^{\ast}_L(1,2)_{1/2,-4},e^{\ast}_R(1,1)_{1,2}$\\ \hline
   $21^{(+,+)}\supset L^*_{21} (1,2)_{1/2,-4}^{(-,-)}, N^*_{21} (1,1)_{0,-10}^{(+,+)}$
      &$21^{(-,-)}$
         &$l^{\ast}_L(1,2)_{1/2,-4},\nu^{\ast}_R(1,1)_{0,-10}$\\ \hline
  \end{tabular}
 \caption{Representation of bulk fermions, 
 the corresponding mirror fermions and SM fermions per a generation. 
 $R$ in $R^{(+,+)}$ means an $SU(6)$ representation of the bulk fermion. 
 $r_{1,2}$ in $(r_1, r_2)_{a,b}$ are $SU(3), SU(2)$ representations in the SM, respectively. 
 $a,b$ are $U(1)_Y, U(1)_X$ charges. }
 \label{table:representation}
 \end{table}
 
Lagrangian for the fermions is 
 \bea
 \lag_{{\rm matter}}&=&\sum_{a=20,56,15,21} \left[\overline{\Psi}_ai\Gamma^MD_M\Psi_a
                   +\overline{\tilde{\Psi}}_ai\Gamma^MD_M\tilde{\Psi}_a
                       + \left( \frac{ \lambda_a}{\pi R}\overline{\Psi}_a \tilde{\Psi}_a + {\rm h.c.} \right) \right]\non
 &&+\delta(y)\left[\overline{\Psi}_{10}i\Gamma^{\mu}D_{\mu}\Psi_{10}
                      +\overline{\Psi}_{5^{\ast}}i\Gamma^{\mu}D_{\mu}\Psi_{5^{\ast}}
                      +\overline{\Psi}_1i\Gamma^{\mu}D_{\mu}\Psi_1\right.\non
 &&
      +\sqrt{\frac{2}{\pi R}}\left(\overline{q^{\ast}}_L Q^*_{20}
                        +\overline{q}_L Q_{56}
                         +\overline{u^{\ast}}_R U^*_{20}
                        +\overline{d}_R D_{56} \right.\non
 &&
       +\left.\left.\overline{l}^{\ast}_L(L^*_{15} + L^*_{21})
                      +\overline{e}^{\ast}_R E^*_{15}
                      +\overline{\nu^{\ast}}_R N^*_{21} + {\rm h.c.} \right) \right].
 \eea
The first line is lagrangian for the bulk and the corresponding mirror fermions, 
 and the remaining terms are lagrangian localized on $y=0$ boundary. 
Note that the subscript $``a"$ denotes the representations of the bulk and mirror fermions. 
The bulk masses between the bulk and the mirror fermions are normalized by $\pi R$ 
 and expressed by the dimensionless parameter $\lambda_a$. 
The last two lines are mixing mass terms between the bulk fermions and the SM fermions. 
In general, these mixing masses can be free parameters, 
 but we set them to be a common value $\sqrt{2/\pi R}$ 
 since we would like to avoid unnecessary arbitrary parameters 
 in fitting the data of SM fermion masses.  
The five-dimensional gamma matrices $\Gamma^M$ is given by $(\Gamma^{\mu},\Gamma^y)=(\gamma^{\mu},i\gamma^5)$. 
By integrating out $y$-direction, 4D effective Lagrangian from the bulk lagrangian is obtained. 
 \bea
 \mathcal{L}_4 &\supset& \sum_{n = - \infty}^{\infty}
            \Big[ \overline{\Psi}^{(n)} (i \Slash{\partial} - m(q\alpha)) \Psi^{(n)}
            + \overline{\tilde{\Psi}}^{(n)} (i \Slash{\partial} + m(q\alpha)) \tilde{\Psi}^{(n)} \non
 &&
 + \left.\left( \frac{\lambda}{\pi R} \overline{\Psi}^{(n)} \tilde{\Psi}^{(n)}
            + \overline{\psi_{{\rm SM}}} \frac{\kappa_L P_L + \kappa_R P_R}{\pi R} \Psi^{(n)}
            + \mbox{h.c.} \right) \right], 
\label{4dL}
 \eea
 where $\Psi^{(n)}(\tilde{\Psi}^{(n)})$ represents a $n$-th KK mode of bulk (mirror) fermion, 
 and $\psi_{{\rm SM}}$ is a SM fermion. 
$P_{L,R}$ are chiral projection operators and $\kappa_{L,R}$ are some constants. 
$m(q\alpha) = \frac{n + q\alpha}{R}$ denotes the sum of the ordinary KK mass 
 and the electroweak symmetry breaking mass proportional to the Higgs VEV. 
The factor $q$ determined by the representation which the fermion under consideration belongs to. 
The mass spectrum of bulk and mirror fermions is totally given 
 by $m_n^2 = \left(\frac{\lambda}{\pi R}\right)^2 + m(q\alpha)^2$.
Note that the Lagrangian (\ref{4dL}) is illustrated for particular bulk and mirror fermions as an example. 


In order to derive the SM fermion masses, 
 we need the quadratic terms in the effective Lagrangian for the SM fermion. 
 \bea
 {\cal L}_{{\rm SM}} \supset \overline{\psi}_{{\rm SM}} K \psi_{{\rm SM}}
 \eea
with 
 \beq
 \label{eq:K}
 K \equiv \Slash{p} \left( 1 + \frac{\kappa_L P_L + \kappa_R P_R}{\sqrt{x^2 + \lambda^2}} \right)
 \mbox{Re} f(\sqrt{x^2 + \lambda^2}, q \alpha)
    +\frac{i}{\pi R} \mbox{Im} f(\sqrt{x^2 + \lambda^2}, q \alpha) 
 \eeq
where $x \equiv \pi R p$ and  
 \beq
 f(\sqrt{x^2 + \lambda^2}, q \alpha)
 \equiv \sum^{\infty}_{n = - \infty} \frac{1}{\sqrt{x^2 + \lambda^2} + i \pi (n + q \alpha)}
 = \mbox{coth}(\sqrt{x^2 + \lambda^2} + i \pi \alpha).
 \eeq
In deriving ${\cal L}_{{\rm SM}}$, 
 we simply took the large bulk mass limit $\frac{\lambda^2}{(\pi R)^2} \gg p^2$  
 so that the mixings of the SM fermions with non-zero KK modes in the mass eigenstate become negligibly small. 

Integrating out all massive bulk fermions and normalizing the kinetic term to be canonical, 
 we obtain the physical mass for the SM fermions.
 \beq 
 m^a_{{\rm phys}} = \frac{m^a}{\sqrt{Z_L^a Z_R^a}} 
 \simeq m_W e^{- \lambda}~(a=u, d, e, \nu)
 \label{physmass}
 \eeq
where the bare mass and the wave function renormalization factors are
\bea
m^a &=& \frac{1}{\pi R} \mbox{Im} f(\sqrt{x^2 + \lambda^2}, q \alpha) , \\
Z^a_{L, R} &=& 1+ \sum_i \frac{\kappa^i_{L,R}}{\sqrt{x^2 + \lambda_i^2}} 
 \mbox{Re} f(\sqrt{x^2 + \lambda_i^2}, q_i \alpha)
\eea
where the summation in $Z^a_{L, R}$ means that it takes a summation 
 for all the bulk fields contributing to mass $m^a$. 
The explicit expressions are shown below. 

\begin{figure}[h]
  \begin{tabular}{c}
  \begin{minipage}[c]{1.00\hsize}
  \centering
  \includegraphics[keepaspectratio, scale=0.9]{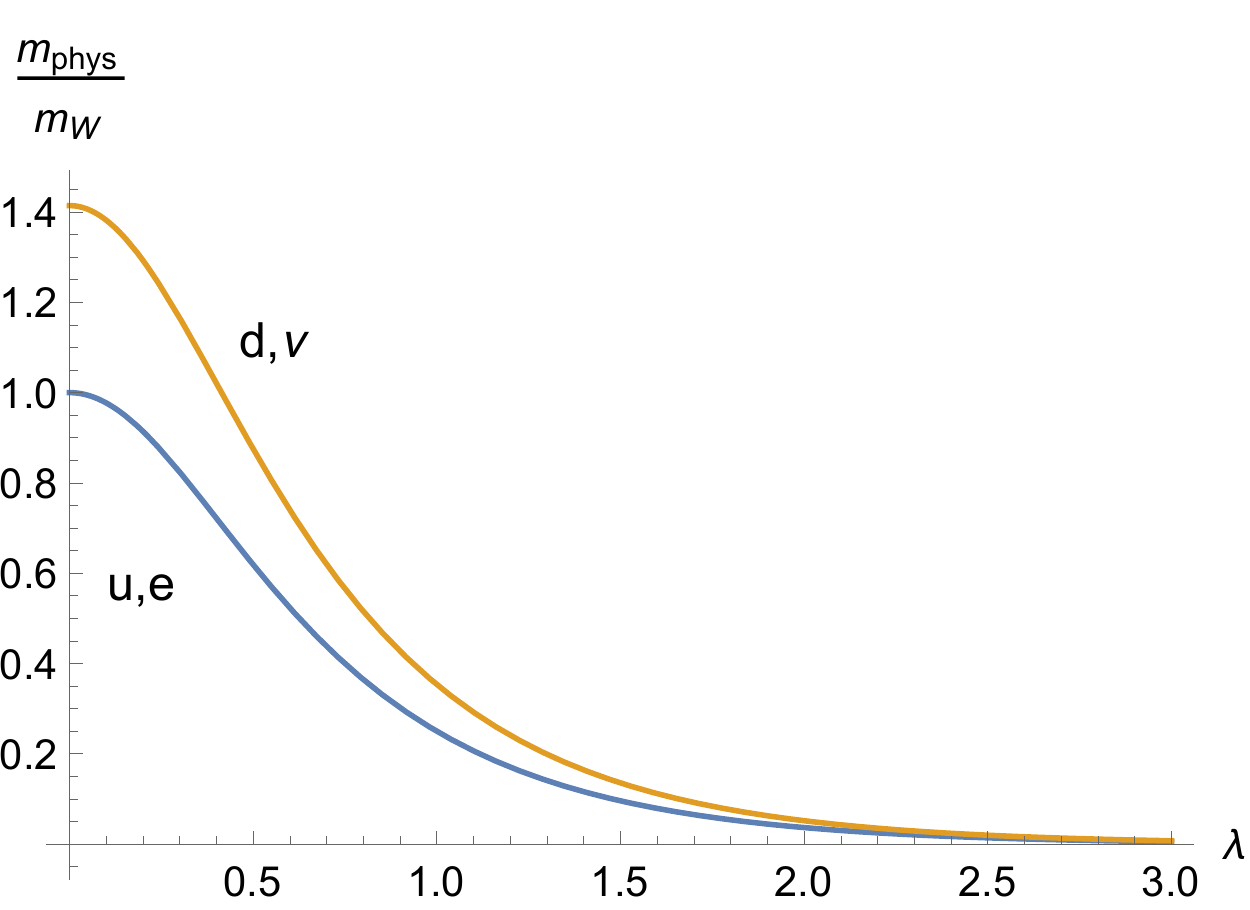}
  \end{minipage}
  \end{tabular}
  \caption{Ratios of the SM fermion mass and the weak boson mass 
  as a function of the bulk mass parameter. 
  The yellow (blue) curve denotes the down-type quark and the neutrino masses 
  (up-type quark and the charged lepton masses).}
  \label{figure:ratio}
 \end{figure}

We consider here ratios of the physical SM fermion mass 
 and the weak boson mass $m_W$ to fit the experimental data. 
\bea
 &&\frac{m_{{\rm phys}}^u}{m_W}
   = \frac{ \left(1 - \coth^2 (\lambda_{20}) \right)}
    {\sqrt{\left( 1 + \frac{1}{\lambda_{20}} \coth(\lambda_{20})
                     + \frac{1}{\lambda_{56}} \coth(\lambda_{56}) \right)
             \left( 1 + \frac{1}{\lambda_{20}} \coth(\lambda_{20}) \right)}}, 
 \\
&& \frac{m_{{\rm phys}}^d}{m_W}
   = \frac{\sqrt{2}\left( 1 - \coth^2 (\lambda_{56}) \right)}
    {\sqrt{\left( 1 + \frac{\epsilon_1^2}{\lambda_{56}} \coth(\lambda_{56}) \right)
    \left( 1 + \frac{\epsilon_2^2}{2\lambda_{56}} \coth(\lambda_{56})
             + \frac{\epsilon_2^2}{2\lambda_{56}} \coth(\lambda_{56})
             + \frac{\epsilon_1^2}{\lambda_{20}} \coth(\lambda_{20}) \right)}}, \non
 \\
&& \frac{m_{{\rm phys}}^e}{m_W}
   = \frac{\left( 1 - \coth^2 (\lambda_{15}) \right)}
     {\sqrt{\left( 1 + \frac{1}{\lambda_{15}} \coth(\lambda_{15})\right)
     \left( 1 + \frac{1}{\lambda_{15}} \coth(\lambda_{15})
              + \frac{1}{\lambda_{21}} \coth(\lambda_{21}) \right)}}, 
 \\
&& \frac{m_{{\rm phys}}^{\nu}}{m_W}
   = \frac{\sqrt{2}\left( 1 - \coth^2(\lambda_{21}) \right)}
     {\sqrt{\left( 1 + \frac{1}{2\lambda_{21}} \coth(\lambda_{21})
                      + \frac{1}{2\lambda_{21}} \coth(\lambda_{21}) \right)
     \left( 1 + \frac{1}{\lambda_{21}} \coth(\lambda_{21})
              +\frac{1}{\lambda_{15}} \coth(\lambda_{15}) \right)}}, \non
 \eea
where $m^{u,d,e,\nu}$ denote up-type quark, down-type quark, 
 charged lepton, and neutrino masses, respectively. 
All these ratios depend on two kinds of bulk mass parameters, 
 but one of them is always dominant to the other one. 
Fig. \ref{figure:ratio} shows the dependence on bulk mass parameter for various mass ratios. 
Note that $\lambda$ in the horizontal axis of the figure means 
 a larger bulk mass parameter of the two kinds: 
 $\lambda = \lambda_{20}, \lambda_{56}, \lambda_{15}, \lambda_{21}$ 
 for up-type quarks, down-type quarks, charged leptons, neutrinos, respectively.
As can be seen from the Figure \ref{figure:ratio}, 
 the masses up to order of the weak boson mass can be realized 
 by choosing an appropriate bulk mass parameter. 
Table \ref{table:bulk mass} summarizes the values of bulk mass parameters 
 reproducing the SM fermion masses except for the top quark. 
It is a very nice feature of models in extra dimensions 
  that the SM fermion mass hierarchy can be obtained 
  by the mild tuning of bulk mass parameters. 
This is because the physical fermion mass has an exponential dependence 
 on the bulk mass parameter as seen from (\ref{physmass}). 
 \begin{table}[h]
  \centering 
   \begin{tabular}{|l|c|c|c|} \hline
    \backslashbox{parameter}{generation} & 1 & 2 & 3 \\ \hline
    $\lambda_{20}$~(up-type~quark) & 5.9 & 2.55 & 0.1 \\ \hline
    $\lambda_{56}$~(down-type~quark) & 5.65 & 4.1 & 1.1 \\ \hline
    $\lambda_{15}$~(charged~lepton) & 6.58 & 3.87 & 2.4 \\ \hline
    $\lambda_{21}$~(neutrino) & 13 & 10 &10  \\ \hline  
  \end{tabular} 
 \caption{Bulk masses fitted by the SM fermion masses except for the top quark mass. 
 }
 \label{table:bulk mass}
 \end{table}
 As for the top quark, even if the vanishing bulk mass parameter is taken, 
  the ratio between top and W-boson masses $m_t/m_W$ is at most unity. 
In order to avoid this situation, the fermion components coupling to top quark on the boundary 
 should be embedded into higher rank representation as in \cite{CCP}.  
We have investigated whether fermions included in three and four rank tensor of $SU(6)$ representations 
 couple to the SM fermions on the $y=0$ boundary, but we could not succeed in finding. 
It might be possible to consider representations on other gauge groups.   
 
\section{Effective potential}
In this section, we calculate the effective potential for the Higgs field 
 and study whether the electroweak symmetry breaking correctly occurs. 
Since the Higgs field is originally a gauge field, 
 the potential is generated at one-loop by Coleman-Weinberg mechanism. 
The potential from the bulk fields is given by
 \bea
 V(\alpha)
             = \sum_n \pm g  \int \frac{d^4p_E}{(2\pi)^4} 
             \log [p_E^2 + m_n^2]
               \equiv g \mathcal{F}^\pm (q\alpha) 
 \eea
 with 
 \beq
 \mathcal{F}^{\pm}(q\alpha) = \pm \sum_n  \int \frac{d^4p_E}{(2\pi)^4}  
 \log [p_E^2+m_n^2],
 \eeq
where overall signs $+(-)$ stand for fermion (boson), respectively. 
$g$ means the spin degrees of freedom of the field running in the loop. 
The loop momentum $p_E$ is taken to be Euclidean. 

For the gauge bosons, bulk fermions and mirror fermions, 
 the mass spectrum is calculated as the following four types 
 depending on the $Z_2$ parity and the bulk mass. 
 \bea
 && m_n^2 = \frac{(n + q \alpha)^2}{R^2}, \non
 && m_n^2 = \frac{(n + 1/2 + q \alpha)^2}{R^2}, \non
 && m_n^2 = \frac{(n + q \alpha)^2}{R^2} + \left( \frac{\lambda}{\pi R} \right)^2, \non
 && m_n^2 = \frac{(n + 1/2 + q \alpha)^2}{R^2} + \left( \frac{\lambda}{\pi R} \right)^2. 
 \eea
Using this information, we obtain the corresponding potentials \cite{CCP}. 
 \bea
 \mathcal{F}^{\pm}(q\alpha) &=&
        \mp \frac{3}{64\pi^6 R^4} \sum_{k=1}^{\infty} \frac{\cos (2\pi q\alpha k)}{k^5}, \non
 \mathcal{F}^{\pm}_{1/2}(q\alpha) &=&
        \mp \frac{3}{64\pi^6 R^4} \sum_{k=1}^{\infty}(-1)^k \frac{\cos (2\pi q \alpha k)}{k^5}, \non
 \mathcal{F}^{\pm}_\lambda (q\alpha)&=&
       \mp \frac{3}{64\pi^6 R^4} \sum_{k=1}^{\infty}
       \frac{\cos(2\pi q \alpha k) e^{-2 k \lambda}} {k^3} 
       \left[\frac{(2\lambda)^3}{3} + \frac{2\lambda}{k}+\frac{1}{k^2}\right],\non
 \mathcal{F}^{\pm}_{1/2\lambda}(q\alpha)&=&
       \mp \frac{3}{64\pi^6 R^4}\sum_{k=1}^{\infty}
       (-1)^k\frac{\cos(2\pi q \alpha k)e^{-2k \lambda}} {k^3}
       \left[\frac{(2\lambda)^3}{3}+\frac{2\lambda}{k}+\frac{1}{k^2}\right].
 \eea
Table \ref{table:potential} lists the various potentials 
 from the gauge field, bulk fermion and mirror fermion contributions.
The coefficients in the potential can be read 
 from the branching rules in the decomposition of the $SU(6)$ representation 
 into $SU(3)_C \times SU(2)_L \times U(1)_Y \times U(1)_X$ representations listed in Appendix A.  

 \begin{table}[h]
  \centering 
   \begin{tabular}{|c||c|} \hline
   bulk+mirror
      &$g=8$      \\ \hline
   $20^{*(-,-)}+20^{*(+,+)}$      
      & $3\mathcal{F}^-_{\lambda}(\alpha)
         +3\mathcal{F}^-_{1/2\lambda}(\alpha)$ \\ \hline
   $56^{(-,+)}+56^{(+,-)}$       
      &$3\mathcal{F}^-_{\lambda}(\alpha)
        +3\mathcal{F}^-_{\lambda}(2\alpha)
        +7\mathcal{F}^-_{1/2\lambda}(\alpha)
        +\mathcal{F}^-_{1/2\lambda}(2\alpha)
        +\mathcal{F}^-_{1/2\lambda}(3\alpha)$ \\ \hline
   $15^{(+,+)} +15^{(-,-)}$        
      &$\mathcal{F}^-_{\lambda}(\alpha)
       +3\mathcal{F}^-_{1/2\lambda}(\alpha)$        \\ \hline
   $21^{(+,+)} +21^{(-,-)}$         
      &$\mathcal{F}^-_{\lambda}(\alpha)
        +\mathcal{F}^-_{\lambda}(2\alpha)
        +3\mathcal{F}^-_{1/2\lambda}(\alpha)$       \\ \hline \hline
   gauge 
       & $g=3$     \\ \hline
   $35^{(+,+)}$       
      &$2\mathcal{F}^+(\alpha)
       +\mathcal{F}^+(2\alpha)
       +6\mathcal{F}^+(\alpha)$   \\ \hline
  \end{tabular}
  \caption{Bulk fermion, mirror fermion and gauge field contributions to Higgs potential.}
  \label{table:potential}
 \end{table}

Next, we have to calculate the Higgs potential from the SM fermion contributions localized at $y=0$ 
using $K$ in eq.(\ref{eq:K}). 
The results are as follows. 
 \bea 
&& V_u =
    -\frac{1}{4\pi^6 R^4}\int dx \,x^3 \non
  && \hspace*{20pt} 
     \times \log\left[\left(1
    +\frac{1}{\sqrt{x^2+\lambda_{20}^2}}\mbox{Re}f(\sqrt{x^2+\lambda_{20}^2},\alpha)
    +\frac{1}{\sqrt{x^2+\lambda_{56}^2}}\mbox{Re}f(\sqrt{x^2+\lambda_{56}^2},\alpha)  
    \right), \right.\non
    && \hspace{20pt}
     \times \left.\left(1
    +\frac{1}{\sqrt{x^2+\lambda_{20}^2}}\mbox{Re}f(\sqrt{x^2+\lambda_{20}^2},\alpha)\right)
    +\left(\frac{1}{x}\mbox{Im}f(\sqrt{x^2+\lambda_{20}^2},\alpha)\right)^2\right], \non
   &&\non
 && V_d = 
    -\frac{1}{4\pi^6 R^4}\int dx\, x^3\log\left[\left(1
    +\frac{1}{\sqrt{x^2+\lambda_{56}^2}}\mbox{Re}f(\sqrt{x^2+\lambda_{56}^2},2\alpha)\right)\right.\non
    && \hspace{20pt}
    \times \left(1
    +\frac{1}{2\sqrt{x^2+\lambda_{56}^2}}\mbox{Re}f(\sqrt{x^2+\lambda_{56}^2},2\alpha)
    +\frac{1}{2\sqrt{x^2+\lambda_{56}^2}}\mbox{Re}f(\sqrt{x^2+\lambda_{56}^2},0)\right.\non
    &&\left. \hspace{20pt}
    +\frac{1}{\sqrt{x^2+\lambda_{20}^2}}\mbox{Re}f(\sqrt{x^2+\lambda_{20}^2},0)\right)
    +\left.\left(\frac{1}{\sqrt{2}x}\mbox{Im}f(\sqrt{x^2+\lambda_{56}^2},2\alpha)\right)^2\right], \non
    &&\non
&& V_e =
    -\frac{1}{4\pi^6 R^4}\int dx\, x^3\log\left[\left(1
    +\frac{1}{\sqrt{x^2+\lambda_{15}^2}}\mbox{Re}f(\sqrt{x^2+\lambda_{15}^2},\alpha))\right)\right.\non
    && \hspace{20pt}
    \times \left(1
    +\frac{1}{\sqrt{x^2+\lambda_{15}^2}}\mbox{Re}f(\sqrt{x^2+\lambda_{15}^2},\alpha)
    +\frac{1}{\sqrt{x^2+\lambda_{21}^2}}\mbox{Re}f(\sqrt{x^2+\lambda_{21}^2},\alpha)\right)\non
    && \hspace{20pt}
    \left.
    +\left(\frac{1}{x}\mbox{Im}f(\sqrt{x^2+\lambda_{15}^2},\alpha)\right)^2\right], \non
    &&\non
&& V_{\nu} =
    -\frac{1}{4\pi^6 R^4}\int dx\, x^3 \non
    && \hspace*{20pt} \times \log\left[\left(1
    +\frac{1}{2\sqrt{x^2+\lambda_{21}^2}}\mbox{Re}f(\sqrt{x^2+\lambda_{21}^2},2\alpha)
    +\frac{1}{2\sqrt{x^2+\lambda_{21}^2}}\mbox{Re}f(\sqrt{x^2+\lambda_{21}^2},0)\right)\right.\non
    &&\hspace{20pt}\times \left(1
    +\frac{1}{\sqrt{x^2+\lambda_{21}^2}}\mbox{Re}f(\sqrt{x^2+\lambda_{21}^2},2\alpha)
    +\frac{1}{\sqrt{x^2+\lambda_{15}^2}}\mbox{Re}f(\sqrt{x^2+\lambda_{15}^2},0)\right)\non
    &&\hspace{20pt}
    +\left.\left(\frac{1}{\sqrt{2}x}\mbox{Im}f(\sqrt{x^2+\lambda_{21}^2},2\alpha)\right)^2\right]. 
 \eea
In calculation of the potential from the both bulk and boundary contributions, 
 we have subtracted the $\alpha$ independent part of the potential 
 since it corresponds to the divergent vacuum energy 
 and is irrelevant to the electroweak symmetry breaking. 

\begin{figure}[h]
  \begin{tabular}{cc}
  \begin{minipage}[c]{0.50\hsize}
  \centering
  \includegraphics[keepaspectratio, scale=0.50]{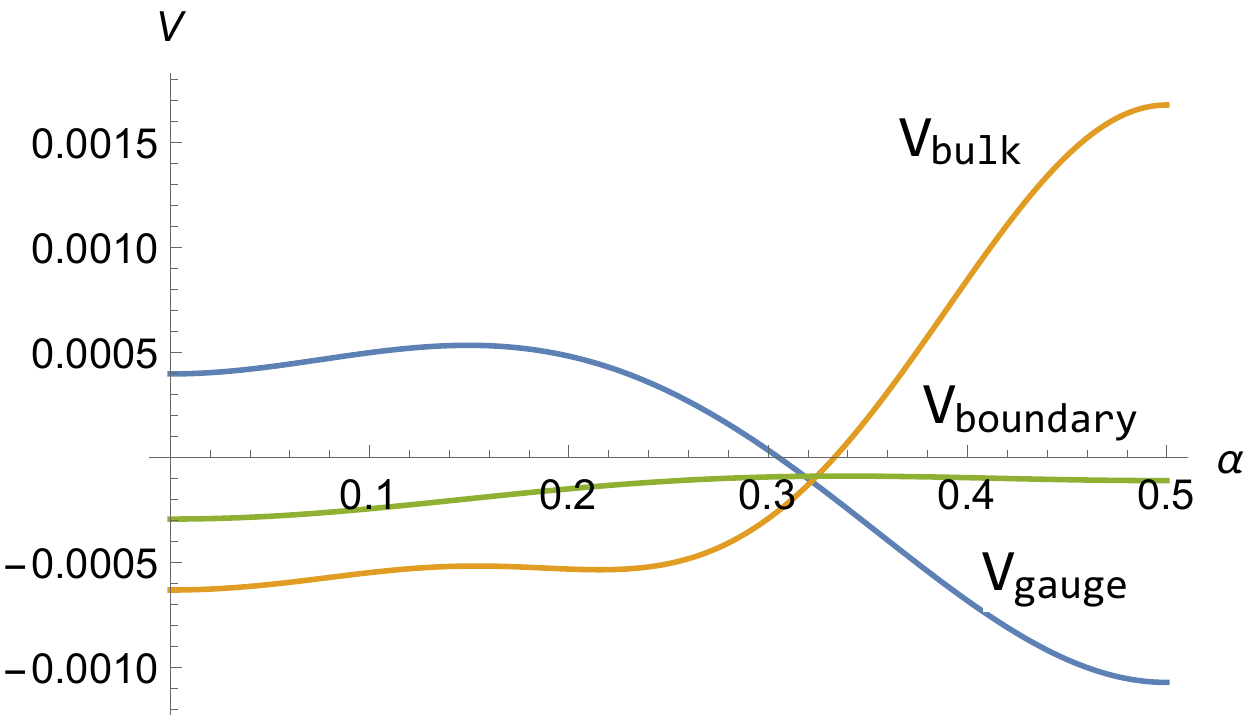}
  \end{minipage}&
  \begin{minipage}[c]{0.50\hsize}
  \centering
  \includegraphics[keepaspectratio, scale=0.50]{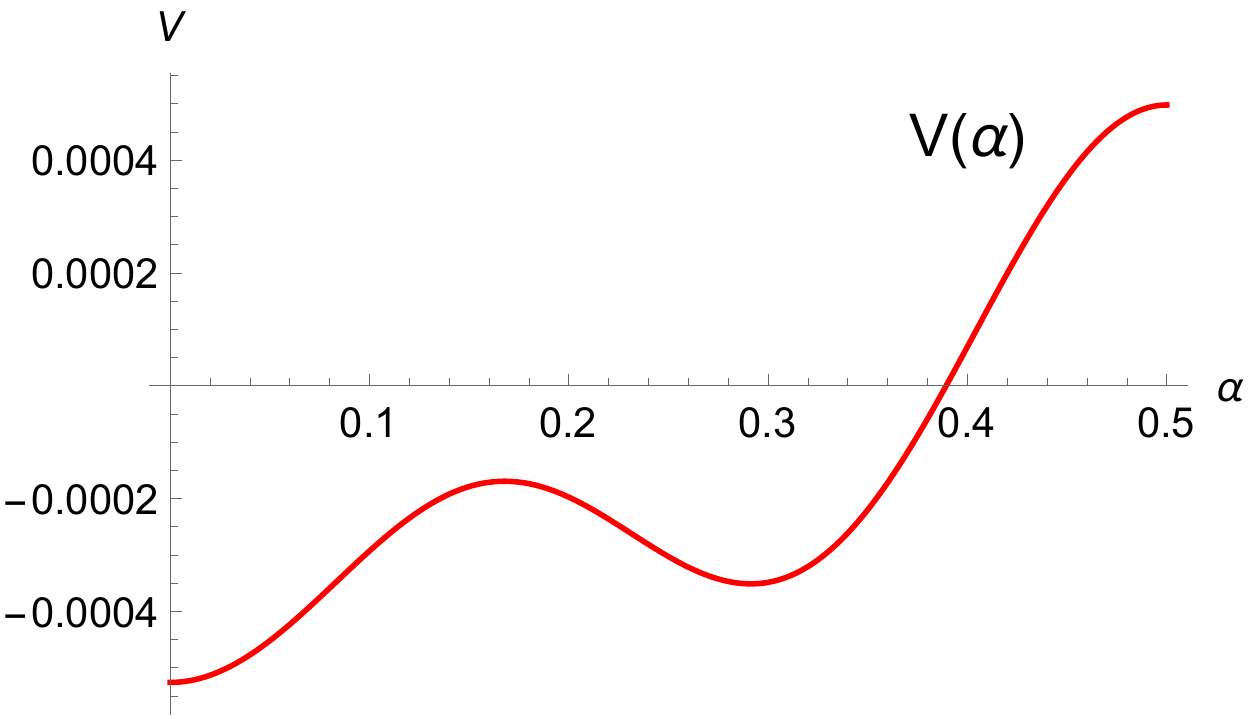}
  \end{minipage}
  \end{tabular}
  \caption{Left: Each contribution of the effective potential; 
  the blue line, the yellow line and green line 
  corresponds to the gauge field, bulk fermion 
  and boundary fermion contributions, respectively. 
  Right: Total Higgs potential.}
  \label{figure:original potential}
 \end{figure}

Total potential is $V(\alpha)=V_{{\rm gauge}}+V_{{\rm bulk}}+V_{{\rm boundary}}$, 
 where $V_{{\rm gauge}}$, $V_{{\rm bulk}}$ and $V_{{\rm boundary}}$ denote 
 the contributions from the gauge field, the bulk fermions and mirror fermions respectively. 
The plots of the potentials are shown in Fig. \ref{figure:original potential}.  
As we can see from Fig. \ref{figure:original potential}, 
 the electroweak symmetry breaking does not occur 
 since the potential minimum is origin.

Therefore, we must add some extra fields to obtain the electroweak symmetry breaking. 
In this paper, we introduce a set of bulk and mirror fermion in {\bf 126} representation of $SU(6)$, 
 which is the fourth rank symmetric tensor. 
The reason why such a bulk fermion with a large dimensional representation is considered is as follows. 
As a generic feature of Higgs potential in GHU, 
 the curvature at the origin of the potential from the gauge field (bulk fermion) contribution is positive (negative) 
 and is likely to make the electroweak symmetry unbroken (broken). 
Furthermore, to realize the electroweak symmetry breaking $SU(2)_L \times U(1)_Y \to U(1)_{{\rm em}}$ in GHU, 
 the Higgs VEV (more precisely, the dimensionless constant Higgs VEV) must be smaller than one, $0 < \alpha < 1$. 
In order to obtain such a small VEV, the field with larger representation is preferable 
 since the periodicity of the potential becomes smaller. 
The additional contribution to Higgs potential is shown in Table \ref{table:potential extra}.

 \begin{table}[h]
  \centering 
   \begin{tabular}{|c||c|} \hline
   extra
      &$g=8$      \\ \hline
   $126^{(+,+)}+126^{(-,-)}$
     &\begin{tabular}{c}       
         $7\mathcal{F}^-_{M}(\alpha)
          +7\mathcal{F}^-_{M}(2\alpha)
           +\mathcal{F}^-_{M}(3\alpha)
            +\mathcal{F}^-_{M}(4\alpha) $ \\
             $+13\mathcal{F}^-_{1/2M}(\alpha)
              +3\mathcal{F}^-_{1/2M}(2\alpha)
               +3\mathcal{F}^-_{1/2M}(3\alpha)$ 
     \end{tabular}\\ \hline
  \end{tabular}
  \caption{The extra bulk fermion contribution to Higgs potential. }
  \label{table:potential extra}
 \end{table}

\begin{figure}[h]
  \begin{tabular}{cc}
  \begin{minipage}[c]{0.50\hsize}
  \centering
  \includegraphics[keepaspectratio, scale=0.5]{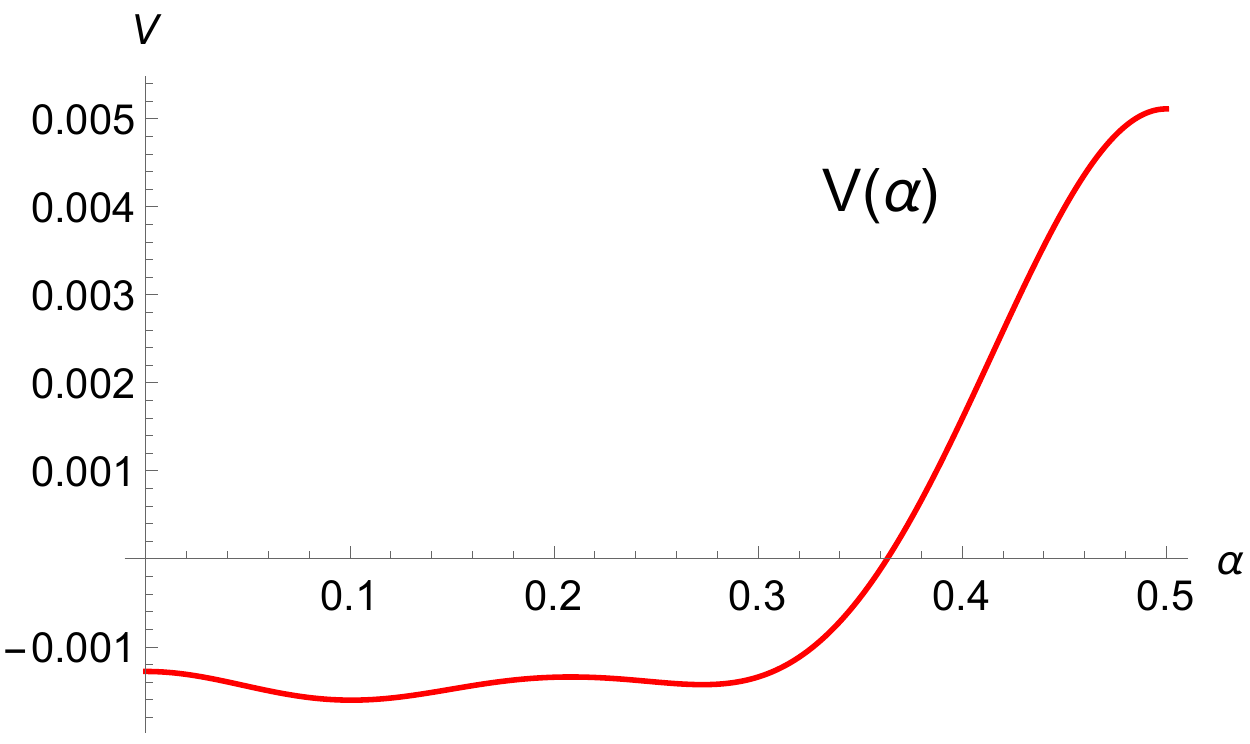}
  \end{minipage}&
  \begin{minipage}[c]{0.50\hsize}
  \centering
  \includegraphics[keepaspectratio, scale=0.5]{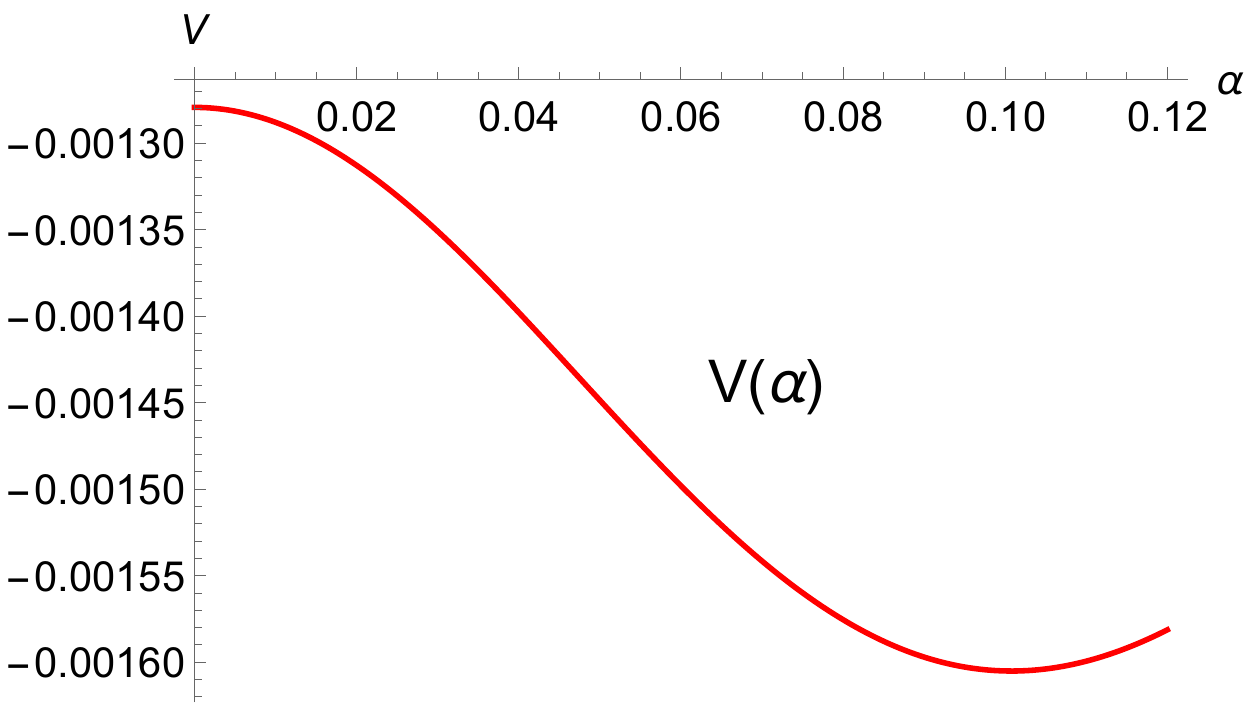}
  \end{minipage}
  \end{tabular}
  \caption{Total potential corrected by adding extra fermions {\bf 126} 
  in the range $0 \leq \alpha \leq 0.5$ (left) and $0 \leq \alpha \leq 0.12$ (right). 
  The bulk mass parameter $\lambda_{126}$ is taken to be 0.5.}
  \label{figure:changed potential}
 \end{figure}

The corrected total potential by adding the contribution from a pair of fermions in ${\bf 126}$ representation is displayed 
 in Fig. \ref{figure:changed potential}, where the bulk mass parameter $\lambda_{126}$ is taken to be 0.5. 
As Fig. \ref{figure:changed potential} shows, the realistic electroweak symmetry breaking is realized. 
In fact, Higgs VEV $\alpha \sim 0.01$ is found from the minimization of the potential. 
Higgs mass can be obtained as a function of the compactification scale. 
We find Higgs mass $m_H \sim 147g_4$ GeV at the compactification scale $1/R \sim 0.8$ TeV. 
$g_4$ is a four-dimensional $SU(2)_L$ gauge coupling obtained 
 from the five-dimensional one $g^2=2 \pi R \, g_4^2$.

\section{Conclusions and discussions}
In this paper, we have considered the fermion mass hierarchy in grand GHU. 
In the grand GHU previously discussed \cite{LM}, 
 a 5D $SU(6)$ GHU with an orbifold $S^1/Z_2$ was considered 
 and all the SM fermions were elegantly embedded into a minimal set of $SU(6)$ bulk multiplets 
 without massless exotic fermions, namely anomaly-free matter content. 
However, the down-type Yukawa couplings and the charged lepton Yukawa couplings were not allowed 
 since the left-handed quark (lepton) doublets and the right-handed down quark (charged lepton) singlets 
 were embedded into different $SU(6)$ multiplets 
 and Yukawa couplings in GHU is generated by the gauge interactions. 
From this observation, the SM fermions were introduced in the $SU(5)$ multiplets 
 on the boundary at $y=0$ in this paper. 
We have also introduced some massive bulk fermions in four types of $SU(6)$ representations 
 and couplings between the SM fermions on the boundary and the bulk fermions.  
By integrating out the massive bulk fermions, the SM fermion masses are generated. 
We have shown that the SM fermion masses except for top quark can be reproduced 
 by mild tuning of bulk masses. 
Furthermore, we have calculated one-loop Higgs potential 
 and found that the electroweak symmetry breaking does not occur unfortunately 
 for the fermion matter content mentioned above. 
To resolve this issue, 
 we have clarified that the electroweak symmetry breaking happened 
 by introducing additional bulk fermions in ${\bf 126}$ representation. 
The SM Higgs boson mass was also obtained.  

In our analysis, 
 Higgs boson mass and the compactification scale are slightly small. 
It might be possible to solve these problems 
 by introducing the localized gauge kinetic terms on the boundary, 
 as discussed in \cite{SSS}. 
These terms are not forbidden by symmetry. 
If we consider the localized gauge kinetic term on the boundary at $y=0$ 
 where the top quark is present, 
 the effects of the localized gauge kinetic term enhance 
 the magnitude of the compactification scale. 
This leads to the enhancement of the Higgs mass. 
Furthermore, top quark mass also enhanced as explained in \cite{SSS}. 
To confirm this expectations, 
 we have to reanalyze the mass spectrum and the mode functions for the gauge fields
 since they are corrected by the presence of the localized gauge kinetic terms. 
This direction is very interesting, but remained for our future study. 

There are issues to be explored in a context of GUT scenario, 
 which are not discussed in this paper. 
First, it is important to study the gauge coupling unification. 
It is well known that the gauge coupling running in (flat) extra dimensions 
 is not logarithmic but power dependence on energy scale \cite{DDG}. 
Therefore, the GUT scale is expected to be very low comparing to the conventional 4D GUT,  
 namely, not far from the compactification scale. 
In this analysis, 
 it is very nontrivial whether the unified $SU(6)$ gauge coupling at the GUT scale is perturbative. 
This is because we have introduced relatively many bulk fields in our model, 
 which might lead to Landau pole below the GUT scale. 
For our model to be a physically meaningful GUT model, this issue must be clarified. 
Second issue to be addressed is proton decay. 
The masses of so-called $X, Y$ gauge bosons are also extremely light comparing to the conventional GUT scale. 
Therefore, proton decays very rapidly and our model is immediately excluded 
 by the constraints from the Super Kamiokande data as it stands. 
Dangerous baryon number violating operators have to be forbidden at tree level 
 by imposing symmetry (see \cite{PDUED} for UED case) in order to ensure the proton stability.  
If $U(1)_X$ is broken to some discrete symmetry and this symmetry plays an role for it, 
 it would be very interesting. 
Then, it is desirable to predict the main decay mode at quantum level. 

These issues are beyond the scope of this paper and also remained for our future work.

\section*{Acknowledgments}
This work is supported in part by JSPS KAKENHI Grant Number JP17K05420 (N.M.).

\appendix
\section{Branching rules of bulk fields}
 In this appendix, several branching rules under the symmetry breaking
 \beq
 SU(6)\rightarrow SU(3)_C\times SU(2)_L\times U(1)_Y\times U(1)_X \non
 \eeq
 are summarized. 
 These branching rules are necessary to search for the fields coupling to the SM fermions on the boundary at $y=0$. 
 They are also useful to compute the one-loop effective Higgs potential. 

 \bea
 \mbox{gauge field}
  \begin{cases}
  35^{(+,+)}=
   &(8,1)_{0,0}^{(+,+)} \oplus (1,3)_{0,0}^{(+,+)} \oplus (1,1)_{0,0}^{(+,+)} \oplus (1,1)_{0,0}^{(+,+)} \oplus (1,1)_{0,0}^{(+,+)} \\
   & \oplus (3,1)_{-1/3,6}^{(-,+)} \oplus (3^{\ast},1)_{1/3,-6}^{(-,+)} \oplus (2,1)_{1/2,6}^{(-,-)} \oplus (2,1)_{-1/2,-6}^{(-,-)}\\
   &\oplus (3,2)_{0,-5/6}^{(+,-)} \oplus (3^{\ast},2)_{0,5/6}^{(+,-)}, \\
 \\
   35^{(-,-)}=
   &(8,1)_{0,0}^{(-,-)} \oplus (1,3)_{0,0}^{(-,-)} \oplus (1,1)_{0,0}^{(-,-)} \oplus (1,1)_{0,0}^{(-,-)} \oplus (1,1)_{0,0}^{(-,-)} \\
   & \oplus (3,1)_{-1/3,6}^{(+,-)} \oplus (3^{\ast},1)_{1/3,-6}^{(+,-)} \oplus (2,1)_{1/2,6}^{(+,+)} \oplus (2,1)_{-1/2,-6}^{(+,+)}\\
   & \oplus (3,2)_{0,-5/6}^{(-,+)} \oplus (3^{\ast},2)_{0,5/6}^{(-,+)}. \\
  \end{cases}
 \eea

 \beq\mbox{bulk fermions}
  \begin{cases}
  15^{(+,+)} = &(3^*,1)_{-2/3,2}^{(+,+)} \oplus (3,2)_{1/6, 2}^{(+,-)} \oplus (1,1)_{1, 2}^{(+,+)} \\
             &\oplus (3, 1)_{-1/3, -4}^{(-,+)} \oplus (1,2)_{1/2, -4}^{(-,-)}, \\
\\
  20^{*(-,-)}=&(3, 2)_{1/6, -3}^{(+,-)} \oplus (3^*, 1)_{-2/3, -3}^{(+,+)} \oplus (1,1)_{1, -3}^{(+,+)}\\
                     &\oplus (3^*, 2)_{-1/6, 3}^{(-,-)} \oplus (3, 1)_{2/3, 3}^{(-,+)} \oplus (1,1)_{-1, 3}^{(-,+)}, \\

\\
  21^{(+,+)} = 
             &(6, 1)_{-1/3, 2}^{(+,+)} \oplus (3, 2)_{1/6, 2}^{(+,-)} \oplus (1,3^*)_{1, 2}^{(+,+)}\\
             &\oplus (3, 1)_{-1/3, -4}^{(-,+)} \oplus (1,2)_{1/2, -4}^{(-,-)} \oplus (1,1)_{0, -10}^{(+,+)}, \\

\\
  56^{(-,+)} = 
                    &(10, 1)_{-1, 3}^{(-,-)} \oplus (6, 2)_{-1/6, 3}^{(-,+)} \oplus (3, 3)_{2/3, 3}^{(-,-)} \\
                     &\oplus (1,4)_{3/2, -3}^{(-,+)} \oplus (6, 1)_{-2/3, -3}^{(+,-)} \oplus (3, 2)_{1/6, -3}^{(+,+)}\\
                      &\oplus (1,3)_{1, -3}^{(+,-)} \oplus (3,1)_{-1/3, -9}^{(-,-)} \oplus (1,2)_{1/2, -9}^{(-,+)} 
                        \oplus (1,1)_{0, -15}^{(+,-)}. \\
 \end{cases}
\eeq

\section{Mass spectrum of Bulk fermions}
In this appendix, the calculations of mass spectrum of bulk fermions 
 are described in detail. 

\subsection{Mode expansion and reflection}
Because of two $Z_2$ parity conditions, 
 the five-dimensional field can be decomposed into four types of KK-modes 
 classified by combination of $Z_2$ eigenvalues $(P,P')$, 
 where the left (right) parity is with respect to $y=0 (\pi R)$. 
The mode functions are listed below.
 \bea
  f_{(+,+)}^{(n)}(y) &=&\frac{1}{\sqrt{2\pi R}}\cos\left(\frac{n}{R}y\right), \non
  f_{(-,-)}^{(n)}(y) &=&\frac{1}{\sqrt{2\pi R}}\sin\left(\frac{n}{R}y\right), \non
  f_{(+,-)}^{(n)}(y) &=&\frac{1}{\sqrt{2\pi R}}\cos\left(\frac{n+1/2}{R}y\right), \non
  f_{(-,+)}^{(n)}(y) &=&\frac{1}{\sqrt{2\pi R}}\sin\left(\frac{n+1/2}{R}y\right). 
 \eea
It is convenient to define the following reflection properties for KK-modes.
 \beq
 \begin{cases}
  \psi^{(-n)}=\psi^{(n)}\hspace{49pt}\hbox{for}~(+,+), \\
  \psi^{(-n)}=-\psi^{(n)}\hspace{39.5pt}\hbox{for}~(-,-), \\
  \psi^{(-n-1)}=\psi^{(n)}\hspace{38pt}\hbox{for}~(+,-), \\
  \psi^{(-n-1)}=-\psi^{(n)}\hspace{28.5pt}\hbox{for}~(-,+). \\
  \end{cases}
 \eeq
Utilizing these reflection properties,  
 the five-dimensional field $\Psi(x,y)$ is expanded 
 in terms of mode function $f(y)$ and four-dimensional field $\psi(x)$ as follows.

As an example, the KK decomposition of the field with $(+,+)$ parity is discussed in detail.  
 \bea
 \Psi(x, y)_{(+,+)}
 &=& \frac{1}{\sqrt{2\pi R}}\left[\psi^{(0)}(x)
            + \sqrt{2}\sum_{n=1}^{\infty}\cos \left(\frac{n}{R}y\right)\psi^{(n)}(x) \right]\non
 &=&\frac{1}{\sqrt{2\pi R}}\psi^{(0)}(x) + \sum_{n=1}^{\infty}\frac{1}{\sqrt{2}}f^{(n)}_{(+,+)}(y) \psi^{(n)}(x)
     + \sum_{n=1}^{\infty}\frac{1}{\sqrt{2}}f^{(n)}_{(+,+)}(y) \psi^{(n)}(x) \non
 &=&\frac{1}{\sqrt{2\pi R}}\psi^{(0)}(x)
     +\sum_{n=1}^{\infty}\frac{1}{\sqrt{2}}f^{(n)}_{(+,+)}\psi^{(n)}(x)
     +\sum_{n=-\infty}^{-1}\frac{1}{\sqrt{2}}f^{(-n)}_{ (+,+)}(y) \psi^{(-n)}(x) \non
 &=&\frac{1}{\sqrt{2\pi R}}\psi^{(0)}(x) + \sum_{n=1}^{\infty}\frac{1}{\sqrt{2}}f^{(n)}_{(+,+)}(y) \psi^{(n)}(x)
     +\sum_{n=-\infty}^{-1}\frac{1}{\sqrt{2}}f^{(n)}_{(+,+)}(y) \psi^{(n)}(x) \non
 &=&\sum_n \eta_n f^{(n)}_{(+,+)}(y) \psi^{(n)}(x)
 \eea
 where $\eta_n=
 \begin{cases}
 1\hspace{18pt}\mbox{for}\,n=0, \\
 \frac{1}{\sqrt{2}}\hspace{10pt}\mbox{for}\,n\neq0. \\
 \end{cases}$
\\

Other types of fields can be also decomposed in a similar way and we obtain the results as 
 \bea
 \Psi(x, y)_{(-,-)}
 &=&\sum_n \eta_n f^{(n)}_{(-,-)}(y) \psi^{(n)}(x), \non
 \Psi(x,y)_{(+,-)} 
 &=&\sum_n \frac{1}{\sqrt{2}} f^{(n)}_{(+,-)}(y) \psi^{(n)}(x), \non
 \Psi(x, y)_{(-,+)} 
 &=&\sum_n \frac{1}{\sqrt{2}} f^{(n)}_{(-,+)}(y) \psi^{(n)}(x).
 \eea

\subsection{Mass eigenvalues}
We employed four representations of $SU(6)$ as bulk fermion in our model; 
 ${\bf  20}^*$, ${\bf 56}$, ${\bf 15}$ and ${\bf21}$. 
Since all of representations are higher rank representations, 
 it is very nontrivial to find mass eigenvalues after the electroweak symmetry breaking. 
 In this subsection, we describe how the mass eigenvalues are obtained for the above four bulk fields.  
In GHU, the electroweak symmetry breaking masses are generated from the gauge interaction 
 since Higgs field is originated from the fifth component of the gauge field.
 \bea
 \mbox{Tr}\overline{\Psi}i\Gamma^5D_5\Psi = 
   -  \overline{\Psi}_{(-)} D_5 \Psi_{(+)}
                     + \overline{\Psi}_{(+)} D_5 \Psi_{(-)} \nonumber
 \eea
 Turning on the Higgs VEV, we find that the KK masses and the symmetry breaking masses 
  take the following form depending on the tensor structure. 
 \bea
 &\mp&\left( \overline{\Psi}_{(\mp)} D_5 \Psi_{(\pm)} 
  + \overline{\Psi}_{(\pm)} D_5 \Psi_{(\mp)}\right) \non
 &=&
  \begin{cases}
  \mp \overline{\Psi}_{(-)i} \partial_5 \Psi_{(+)i}
    \mp \frac{\alpha}{R}( \overline{\Psi}_{(-)2} \Psi_{(+)6}
      - \overline{\Psi}_{(-)6} \Psi_{(+)2})~({\rm the~first~rank~tensor}), \\
  \mp \overline{\Psi}_{(-)ji} \partial_5 \Psi_{(+)ji}
    \mp2 \frac{\alpha}{R}( \overline{\Psi}_{(-)j2}\Psi_{(+)j6}
      - \overline{\Psi}_{(-)j6}\Psi_{(+)j2})~({\rm the~second~rank~tensor}), \\
  \mp \overline{\Psi}_{(-)ikj} \partial_5\Psi_{(+)ijk}
     \mp3\frac{\alpha}{R}( \overline{\Psi}_{(-)2kj}\Psi_{(+)6jk}
        - \overline{\Psi}_{(-)6kj}\Psi_{(+)2jk})~({\rm the~third~rank~tensor}).
  \end{cases}\non
 \eea
The point is that the coefficients of symmetry breaking mass $\alpha/R$ 
 is determined by the number of rank of the field under consideration. 
Note that the only components 2 and 6 appear in the symmetry breaking terms 
 since the Higgs VEV is supposed to take in $(2,6)$ and $(6,2)$ components in $A_5$. 

In next subsubsections, we briefly discuss how the mass spectrum is derived 
 for each representation.   

\subsubsection{15: the second rank anti-symmetric tensor}

The ${\bf 15}$ representation is the second rank anti-symmetric tensor of $SU(6)$. 
The components after the decomposition into $SU(3)_C \times SU(2)_L \times U(1)_Y \times U(1)_X$ 
and the corresponding parity and reflection are summarized in Table \ref{table:bulk parity 15}. 
The blanks in the matrix elements means zero, hereafter. 

 \begin{table}[h]
  \centering 
   \begin{tabular}{|c|c||c|c|} \hline
   $(+,+)$
     &
       &$(+,-)$
         &\\ \hline
   $(1,1)$
      &$E^{*(-n)}_{15(\pm)} = \mp E^{*(n)}_{15(\pm)}$
         &$(3, 2)$
           &$\zeta^{(-n-1)}_{(\pm)} = \mp\zeta^{(n)}_{(\pm)}$\\ \hline
  $(3^*,1)$
     &$\psi^{(-n)}_{(\pm)} = \mp\psi^{(n)}_{(\pm)}$
       &
         &\\ \hline \hline
  $(-,-)$
    &
      &$(-,+)$
         &\\ \hline
  $(1,2)$
     &$L^{*(-n)}_{15(\pm)} = \pm L^{*(n)}_{15(\pm)}$
       &$(3,1)$
          &$\omega^{(-n-1)}_{(\pm)} = \pm\omega^{(n)}_{(\pm)}$\\ \hline 
  \end{tabular} \non
  \caption{Parity and reflection for components of ${\bf 15}$.}
  \label{table:bulk parity 15}
 \end{table}
Making use of the results in the previous subsection B.1., 
 the KK expansion of ${\bf 15}$ is described in a following matrix form.  
 \bea
 \Psi_{(\pm)}=\frac{1}{\sqrt{2}}\sum_{n=-\infty}^{\infty}&& \left(
  \ba{cc|}
  \hspace{10pt} 
     & \eta_n f_{(\mp,\mp)}^{(n)}E_{15(\pm)}^{(n)} \\
  -\eta_n f_{(\mp,\mp)}^{(n)}E_{15(\pm)}^{(n)} 
     &   \\ \hline
  -\frac{1}{\sqrt{2}} f_{(\mp,\pm)}^{(n)}\zeta_{(\pm)}^{(n)1} 
     & -\frac{1}{\sqrt{2}} f_{(\mp,\pm)}^{(n)}\zeta_{(\pm)}^{(n)2} \\
  -\frac{1}{\sqrt{2}} f_{(\mp,\pm)}^{(n)}\zeta_{(\pm)}^{(n)3} 
     & -\frac{1}{\sqrt{2}} f_{(\mp,\pm)}^{(n)}\zeta_{(\pm)}^{(n)4} \\
  -\frac{1}{\sqrt{2}} f_{(\mp,\pm)}^{(n)}\zeta_{(\pm)}^{(n)5} 
     & -\frac{1}{\sqrt{2}} f_{(\mp,\pm)}^{(n)}\zeta_{(\pm)}^{(n)6}  \\ \hline
  \mp\eta_n f_{(\pm,\pm)}^{(n)}L_{15(\pm)}^{(n)1} 
     & \mp\eta_n f_{(\pm,\pm)}^{(n)}L_{15(\pm)}^{(n)2}  \\
 \ea\right. \non
 &&\left. 
  \ba{ccc|c}
  \frac{1}{\sqrt{2}} f_{(\mp,\pm)}^{(n)}\zeta_{(\pm)}^{(n)1} 
    & \frac{1}{\sqrt{2}} f_{(\mp,\pm)}^{(n)}\zeta_{(\pm)}^{(n)3} 
      & \frac{1}{\sqrt{2}} f_{(\mp,\pm)}^{(n)}\zeta_{(\pm)}^{(n)5} 
        & \pm\eta_n f_{(\pm,\pm)}^{(n)}L_{15(\pm)}^{(n)1} \\
  \frac{1}{\sqrt{2}} f_{(\mp,\pm)}^{(n)}\zeta_{(\pm)}^{(n)2} 
    & \frac{1}{\sqrt{2}} f_{(\mp,\pm)}^{(n)}\zeta_{(\pm)}^{(n)4} 
      & \frac{1}{\sqrt{2}} f_{(\mp,\pm)}^{(n)}\zeta_{(\pm)}^{(n)6} 
        & \pm\eta_n f_{(\pm,\pm)}^{(n)}L_{15(\pm)}^{(n)2} \\ \hline
    & \eta_n f_{(\mp,\mp)}^{(n)}\psi_{(\pm)}^{(n)1} 
       & \eta_n f_{(\mp,\mp)}^{(n)}\psi_{(\pm)}^{(n)2} 
          & \pm\frac{1}{\sqrt{2}} f_{(\pm,\mp)}^{(n)}\omega_{(\pm)}^{(n)1} \\
  -\eta_n f_{(\mp,\mp)}^{(n)}\psi_{(\pm)}^{(n)1} 
    & 
      & \sqrt{2}\eta_n f_{(\mp,\mp)}^{(n)}\psi_{(\pm)}^{(n)3} 
        & \pm\frac{1}{\sqrt{2}} f_{(\pm,\mp)}^{(n)}\omega_{(\mp)}^{(n)1} \\
  -\eta_n f_{(\mp,\mp)}^{(n)}\psi_{(\pm)}^{(n)2} 
    &- \pm\eta_n f_{(\mp,\mp)}^{(n)}\psi_{(\pm)}^{(n)3} 
      & 
        & \pm\frac{1}{\sqrt{2}} f_{(\pm,\mp)}^{(n)}\omega_{(\pm)}^{(n)3} \\ \hline
  \mp\frac{1}{\sqrt{2}} f_{(\pm,\mp)}^{(n)}\omega_{(\pm)}^{(n)1} 
    & \mp\frac{1}{\sqrt{2}} f_{(\pm,\mp)}^{(n)}\omega_{(\pm)}^{(n)2} 
      & \mp\frac{1}{\sqrt{2}} f_{(\pm,\mp)}^{(n)}\omega_{(\pm)}^{(n)3} 
        &   \\
  \ea \right). \no
 \eea
Substituting this expansion into the mass term and diagonalizing it, 
 we find mass spectrum
 \bea
 \lag_4 \supset
  -\sum_{n=-\infty}^{\infty}\left[\frac{n+\alpha}{R} \overline{\Psi}^{(n)1}_{(\pm)}\Psi^{(n)1}_{(\pm)}
      +\sum_{i=2}^{4}\frac{n+1/2+\alpha}{R} \overline{\Psi}^{(n)i}_{(\pm)}\Psi^{(n)i}_{(\pm)}\right]\non
  -\sum_{n=1}^{\infty}\left[\sum_{i=5}^{8}\frac{n}{R} \overline{\Psi}^{(n)i}_{(\pm)}\Psi^{(n)i}_{(\pm)}
    +\sum_{i=9}^{11}\frac{n+1/2}{R} \overline{\Psi}^{(n)i}_{(\pm)}\Psi^{i(n)}_{(\pm)}\right]
 \eea
and the corresponding mass eigenstates are given by
 \bea
 &&\Psi^{(n)1}_{(\pm)}=\eta_n\left(L^{(n)1}_{15(\pm)}+E^{(n)}_{15(\pm)}\right), \quad 
   \Psi^{(n)\{2,3,4\}}_{(\pm)}=\frac{1}{\sqrt{2}}\left(\omega^{(n)\{1,2,3\}}_{(\pm)}
                                                                       -\zeta^{(n)\{1,2,3\}}_{(\pm)}\right), \non
 &&\Psi^{(n)\{5,6,7\}}_{(\pm)}=\psi^{(n)\{1,2,3\}}_{(\pm)}, \quad
  \Psi^{(n)8}_{(\pm)}=L^{(n)2}_{15(\pm)}, \quad 
 \Psi^{(n)\{9,10,11\}}_{(\pm)}=\zeta^{(n)\{4,5,6\}}_{(\pm)}. 
 \eea
 
\subsubsection{21: the second rank symmetric tensor}

The ${\bf 21}$ representation is the second rank symmetric tensor of $SU(6)$. 
The components after the decomposition into $SU(3)_C \times SU(2)_L \times U(1)_Y \times U(1)_X$ 
and the corresponding parity and reflection are summarized in Table \ref{table:bulk parity 21}. 

 \begin{table}[h]
  \centering 
   \begin{tabular}{|c|c||c|c|} \hline
   $(+,+)$
    &
     &$(+,-)$
      &\\ \hline
   $(1,3)$
    &$\phi^{(-n)}_{(\pm)}=\mp\phi^{(n)}_{(\pm)}$
     &$(3,2)$
      &$\zeta^{(-n-1)}_{(\pm)}=\mp\zeta^{(n)}_{(\pm)}$\\ \hline
   $(6,1)$
    &$\psi^{(-n)}_{(\pm)}=\mp\psi^{(n)}_{(\pm)}$
     &
      &\\ \hline
   $(1,1)$
    &$N^{*(-n)}_{21(\pm)}=\mp N^{*(n)}_{21(\pm)}$
     &
      &\\ \hline \hline
   $(-,-)$
    &
     &$(-,+)$
      &\\ \hline
   $(1,2)$
    &$L^{*(-n)}_{21(\pm)}=\pm L^{*(n)}_{21(\pm)}$
     &$(3,1)$
      &$\omega^{(-n-1)}_{(\pm)}=\pm  \omega^{(n)}_{(\pm)}$\\ \hline 
  \end{tabular} \non
  \caption{Parity and reflection for components of ${\bf 21}$.}
  \label{table:bulk parity 21}
 \end{table}
KK expansion and diagonalization of mass matrix can proceed 
 similarly as ${\bf 15}$ representation in the previous subsubsection. 
 \bea
 \Psi_{(\pm)}&=&\frac{1}{\sqrt{2}}\sum_{n=-\infty}^{\infty} \left(
   \ba{cc|}
   \sqrt{2}\eta_n f_{(\mp,\mp)}^{(n)}\phi_{(\pm)}^{(n)1} 
         & \eta_n f_{(\mp,\mp)}^{(n)}\phi_{(\pm)}^{(n)2} \\
   \eta_n f_{(\mp,\mp)}^{(n)}\phi_{(\pm)}^{(n)2} 
         & \sqrt{2}\eta_n f_{(\mp,\mp)}^{(n)}\phi_{(\pm)}^{(n)3}  \\ \hline
   \frac{1}{\sqrt{2}} f_{(\mp,\pm)}^{(n)}\zeta_{(\pm)}^{(n)1} 
         & \frac{1}{\sqrt{2}} f_{(\mp,\pm)}^{(n)}\zeta_{(\pm)}^{(n)2} \\
   \frac{1}{\sqrt{2}} f_{(\mp,\pm)}^{(n)}\zeta_{(\pm)}^{(n)3} 
         & \frac{1}{\sqrt{2}} f_{(\mp,\pm)}^{(n)}\zeta_{(\pm)}^{(n)4} \\
  \frac{1}{\sqrt{2}} f_{(\mp,\pm)}^{(n)}\zeta_{(\mp)}^{(n)5} 
         & \frac{1}{\sqrt{2}} f_{(\mp,\pm)}^{(n)}\zeta_{(\mp)}^{(n)6}  \\ \hline
  \pm\eta_n f_{(\pm,\pm)}^{(n)}L_{21(\pm)}^{(n)1} 
         &\pm \eta_n f_{(\pm,\pm)}^{(n)}L_{21(\pm)}^{(n)2} \\
  \ea  \right.\non
  &&\left. \ba{ccc|c}
   \frac{1}{\sqrt{2}} f_{(\mp,\pm)}^{(n)}\zeta_{(\pm)}^{(n)1}
      & \frac{1}{\sqrt{2}} f_{(\mp,\pm)}^{(n)}\zeta_{(\pm)}^{(n)3}
         & \frac{1}{\sqrt{2}} f_{(\mp,\pm)}^{(n)}\zeta_{(\pm)}^{(n)5} 
            & \pm\eta_n f_{(\pm,\pm)}^{(n)}L_{21(\pm)}^{(n)1} \\
   \frac{1}{\sqrt{2}} f_{(\mp,\pm)}^{(n)}\zeta_{(\pm)}^{(n)2} 
      & \frac{1}{\sqrt{2}} f_{(\mp,\pm)}^{(n)}\zeta_{(\pm)}^{(n)4}
         & \frac{1}{\sqrt{2}} f_{(\mp,\pm)}^{(n)}\zeta_{(\pm)}^{(n)6} 
            & \pm\eta_n f_{(\pm,\pm)}^{(n)}L_{21(\pm)}^{(n)2} \\ \hline
   \sqrt{2}\eta_n f_{(\mp,\mp)}^{(n)}\psi_{(\pm)}^{(n)1}  
      & \eta_n f_{(\mp,\mp)}^{(n)}\psi_{(\pm)}^{(n)2} 
         &\eta_n f_{(\mp,\mp)}^{(n)}\psi_{(\pm)}^{(n)4} 
           & \pm\frac{1}{\sqrt{2}} f_{(\pm,\mp)}^{(n)}\omega_{(\pm)}^{(n)1} \\
   \eta_n f_{(\mp,\mp)}^{(n)}\psi_{(\pm)}^{(n)2} 
      &\sqrt{2}\eta_n f_{(\mp,\mp)}^{(n)}\psi_{(\pm)}^{(n)3} 
         & \sqrt{2}\eta_n f_{(\mp,\mp)}^{(n)}\psi_{(\pm)}^{(n)5} 
           & \pm\frac{1}{\sqrt{2}} f_{(\pm,\mp)}^{(n)}\omega_{(\pm)}^{(n)2} \\
  \eta_n f_{(\mp,\mp)}^{(n)}\psi_{(\pm)}^{(n)4}
      & \eta_n f_{(\mp,\mp)}^{(n)}\psi_{(\pm)}^{(n)5}
        & \sqrt{2}\eta_n f_{(\mp,\mp)}^{(n)}\psi_{(\pm)}^{(n)6} 
           &\pm \frac{1}{\sqrt{2}} f_{(\pm,\mp)}^{(n)}\omega_{(\pm)}^{(n)3} \\ \hline
  \pm \frac{1}{\sqrt{2}} f_{(\pm,\mp)}^{(n)}\omega_{(\pm)}^{(n)1} 
      &\pm \frac{1}{\sqrt{2}} f_{(\pm,\mp)}^{(n)}\omega_{(\pm)}^{(n)2}
        & \pm\frac{1}{\sqrt{2}} f_{(\pm,\mp)}^{(n)}\omega_{(\pm)}^{(n)3}
           & \sqrt{2}\eta_nf_{(\mp,\mp)}^{(\mp)}N_{21(\pm)}^{(n)} \\
  \ea \right). \no
 \eea
The diagonalized mass terms are
 \bea
 \lag_4 &\supset&
  -\sum_{n=-\infty}^{\infty}\left[\frac{n+\alpha}{R} \overline{\Psi}^{(n)1}_{(\pm)}\Psi^{(n)1}_{(\pm)}
    +\frac{n+2\alpha}{R} \overline{\Psi}^{(n)2}_{(\pm)}\Psi^2_{(n)(\pm)}
      +\sum_{i=3,4,5}\frac{n+1/2+\alpha}{R} \overline{\Psi}^{(n)i}_{(\pm)}\Psi^{(n)i}_{(\pm)}\right]\non
&&  -\sum_{n=1}^{\infty}\left[\sum_{i=6}^{13}\frac{n}{R} \overline{\Psi}^{(n)i}_{(\pm)}\Psi^{(n)i}_{(\pm)}
    +\sum_{i=14}^{16}\frac{n+1/2}{R} \overline{\Psi}^{(n)i}_{(\pm)}\Psi^{i(n)}_{(\pm)}\right]
 \eea
and the corresponding mass eigenstates are given
 \bea
 &&\Psi^{(n)1}_{(\pm)}=\eta_n\left(\phi^{(n)2}_{(\pm)}-L^{(n)1}_{21(\pm)}\right), \quad
   \Psi^{(n)2}_{(\pm)}=\eta_n\left(L^{(n)2}_{21(\pm)}
                                           -\frac{1}{\sqrt{2}}\left(\phi^{(n)3}_{(\pm)}
                                                  -N^{(n)}_{21(\pm)}\right)\right), \non
 &&\Psi^{(n)\{3,4,5\}}_{(\pm)}=\frac{1}{\sqrt{2}}\left(\zeta^{(n)\{2,4,6\}}_{(\pm)}
                                               -\omega^{(n)\{1,2,3\}}_{(\pm)}\right), \quad
  \Psi^{(n)6}_{(\pm)}=\frac{1}{\sqrt{2}}\left(\phi^{(n)3}_{(\pm)}+N^{(n)}_{21(\pm)}\right), \non
 && \Psi^{(n)7}_{(\pm)}=\phi^{(n)1}_{(\pm)}, \quad
\Psi^{(n)\{8,9,10,11,12,13\}}_{(\pm)}=\psi^{(n)\{1,2,3,4,5,6\}}, \quad
 \Psi^{(n)\{1,4,15,16\}}_{(\pm)}=\zeta^{(n)\{1,3,5\}}_{(\pm)}. 
 \eea

\subsubsection{20$^*$: the third rank anti-symmetric tensor}
The ${\bf 20}^*$ representation is the third rank anti-symmetric tensor of $SU(6)$. 
The components after the decomposition into $SU(3)_C \times SU(2)_L \times U(1)_Y \times U(1)_X$ 
and the corresponding parity and reflection are summarized in Table \ref{table:bulk parity 20}. 

 \begin{table}[h]
  \centering 
   \begin{tabular}{|c|c||c|c|} \hline
    $(+,+)$
      &
         &$(+,-)$
           &\\ \hline
    $(3^*, 1)$
      &$U^{*(-n)}_{20(\pm)}=\mp U^{*(n)}_{20(\pm)}$
        &$(3, 2)$
          &$\sigma^{(-n-1)}_{(\pm)}=\mp\sigma^{(n)}_{(\pm)}$\\ \hline
    $(1,1)$
     &$\tau^{(-n)}_{(\pm)}=\mp\tau^{(n)}_{(\pm)}$
      &
        &\\ \hline \hline
    $(-,-)$
      &
        &$(-,+)$
          &\\ \hline
    $(3^*, 2)$
      &$Q^{*(-n)}_{20(\pm)}=\pm Q^{*(n)}_{20(\pm)}$
        &$(3, 1)$
         &$\omega^{(-n-1)}_{(\pm)}=\pm\omega^{(n)}_{(\pm)}$\\ \hline
      &
        &$(1,1)$
          &$\zeta^{(-n-1)}_{(\pm)}=\pm\zeta^{(n)}_{(\pm)}$\\ \hline
  \end{tabular} \non
  \caption{Parity and reflection for components of ${\bf 20}^*$. 
  }
  \label{table:bulk parity 20}
 \end{table}
It is straightforward to extend the KK expansion to the third rank tensor case, 
 but takes a more complicated form. 
 \bea
  (\Psi_{(\pm)1})_{jk}=\Psi_{(\pm)1jk}=\frac{1}{\sqrt{6}}\sum_{n=-\infty}^{\infty} \left( 
   \ba{cc|}
   \hspace{50pt}
      & \\
      &\\ \hline
      & \mp\frac{1}{\sqrt{2}} f_{(\pm,\mp)}^{(n)}\omega_{(\pm)}^{(n)1} \\
      & \mp\frac{1}{\sqrt{2}} f_{(\pm,\mp)}^{(n)}\omega_{(\pm)}^{(n)2}\\
      & \mp\frac{1}{\sqrt{2}} f_{(\pm,\mp)}^{(n)}\omega_{(\pm)}^{(n)3}  \\ \hline
      &  -\eta_n f_{(\mp,\mp)}^{(n)}\tau_{(\pm)}^{(n)} 
  \ea \right.  \non 
  \left. 
  \ba{ccc|c}
    &
      &
        &\\
  \pm\frac{1}{\sqrt{2}} f_{(\pm,\mp)}^{(n)}\omega_{(\pm)}^{(n)1} 
    & \pm\frac{1}{\sqrt{2}} f_{(\pm,\mp)}^{(n)}\omega_{(\pm)}^{(n)2}
      & \pm\frac{1}{\sqrt{2}} f_{(\pm,\mp)}^{(n)}\omega_{(\pm)}^{(n)3} 
        & \eta_n f_{(\mp,\mp)}^{(n)}\tau_{(\pm)}^{(n)} \\ \hline
     & \pm\eta_n f_{(\pm,\pm)}^{(n)}Q_{20(\pm)}^{(n)1} 
      &\pm\eta_n f_{(\pm,\pm)}^{(n)}Q_{20(\pm)}^{(n)2} 
        & \frac{1}{\sqrt{2}} f_{(\mp,\pm)}^{(n)}\sigma_{(\pm)}^{(n)1} \\
  \mp\eta_n f_{(\pm,\pm)}^{(n)}Q_{20(\pm)}^{(n)1} 
    &
      & \pm\eta_n f_{(\pm,\pm)}^{(n)}Q_{20(\pm)}^{(n)3} 
        & \frac{1}{\sqrt{2}} f_{(\mp,\pm)}^{(n)}\sigma_{(\pm)}^{(n)3} \\
  \mp\eta_n f_{(\pm,\pm)}^{(n)}Q_{20(\pm)}^{(n)2} 
    & \mp\eta_n f_{(\pm,\pm)}^{(n)}Q_{20(\pm)}^{(n)3}
      &  
        & \frac{1}{\sqrt{2}} f_{(\mp,\pm)}^{(n)}\sigma_{(\pm)}^{(n)5} \\ \hline
  -\frac{1}{\sqrt{2}} f_{(\mp,\pm)}^{(n)}\sigma_{(\pm)}^{(n)1} 
    & -\frac{1}{\sqrt{2}} f_{(\mp,\pm)}^{(n)}\sigma_{(\pm)}^{(n)3}  
      & -\frac{1}{\sqrt{2}} f_{(\mp,\pm)}^{(n)}\sigma_{(\pm)}^{(n)5} 
        & \\
  \ea \right), \no
 \eea


 \bea
 (\Psi_{(\pm)2})_{jk}=\Psi_{(\pm)2jk}=\frac{1}{\sqrt{6}}\sum_{n=-\infty}^{\infty} \left(
  \ba{cc|}
    & \hspace{50pt}\\
    &   \\ \hline
  \mp\frac{1}{\sqrt{2}} f_{(\pm,\mp)}^{(n)}\omega_{(\pm)}^{(n)1} 
    & \\
  \mp\frac{1}{\sqrt{2}} f_{(\pm,\mp)}^{(n)}\omega_{(\pm)}^{(n)2}
    & \\
  \mp\frac{1}{\sqrt{2}} f_{(\pm,\mp)}^{(n)}\omega_{(\pm)}^{(n)3}
    & \\ \hline
  -\eta_n f_{(\mp,\mp)}^{(n)}\tau_{(\pm)}^{(n)} 
    &   \\
  \ea \right. \non
  \left. 
  \ba{ccc|c}
  \pm\frac{1}{\sqrt{2}} f_{(\pm,\mp)}^{(n)}\omega_{(\pm)}^{(n)1} 
    & \pm\frac{1}{\sqrt{2}} f_{(\pm,\mp)}^{(n)}\omega_{(\pm)}^{(n)2}
      & \pm\frac{1}{\sqrt{2}} f_{(\pm,\mp)}^{(n)}\omega_{(\pm)}^{(n)3} 
        &\eta_n f_{(\mp,\mp)}^{(n)}\tau_{(\pm)}^{(n)} \\
    &
      & 
        & \\ \hline
    & \pm\eta_n f_{(\pm,\pm)}^{(n)}Q_{20(\pm)}^{(n)4} 
      &\pm\eta_n f_{(\pm,\pm)}^{(n)}Q_{20(\pm)}^{(n)5} 
       & \frac{1}{\sqrt{2}} f_{(\mp,\pm)}^{(n)}\sigma_{(\pm)}^{(n)2} \\
  \mp\eta_n f_{(\pm,\pm)}^{(n)}Q_{20(\pm)}^{(n)4} 
    & 
      & \pm\eta_n f_{(\pm,\pm)}^{(n)}Q_{20(\pm)}^{(n)6} 
        & \frac{1}{\sqrt{2}} f_{(\mp,\pm)}^{(n)}\sigma_{(\pm)}^{(n)4} \\
  \mp\eta_n f_{(\pm,\pm)}^{(n)}Q_{20(\pm)}^{(n)5} 
    & \mp\eta_n f_{(\pm,\pm)}^{(n)}Q_{20(\pm)}^{(n)6}
     & 
       & \frac{1}{\sqrt{2}} f_{(\mp,\pm)}^{(n)}\sigma_{(\pm)}^{(n)6} \\ \hline
  -\frac{1}{\sqrt{2}} f_{(\mp,\pm)}^{(n)}\sigma_{(\pm)}^{(n)2} 
    & -\frac{1}{\sqrt{2}} f_{(\mp,\pm)}^{(n)}\sigma_{(\pm)}^{(n)4}  
      & -\frac{1}{\sqrt{2}} f_{(\mp,\pm)}^{(n)}\sigma_{(\pm)}^{(n)6}
       & \\
  \ea\right), \no
 \eea


 \bea
 (\Psi_{(\pm)3})_{jk}=\Psi_{(\pm)3jk}=\frac{1}{\sqrt{6}}\sum_{n=-\infty}^{\infty} 
 \left( 
  \ba{cc|}
    &\pm\frac{1}{\sqrt{2}} f_{(\pm,\mp)}^{(n)}\omega_{(\pm)}^{(n)1}  \\
  \mp\frac{1}{\sqrt{2}} f_{(\pm,\mp)}^{(n)}\omega_{(\pm)}^{(n)1} 
    &     \\ \hline
    &  \\
  \mp\eta_n f_{(\pm,\pm)}^{(n)}Q_{20(\pm)}^{(n)1} 
    & \mp\eta_n f_{(\pm,\pm)}^{(n)}Q_{20(\pm)}^{(n)4}  \\
  \mp\eta_n f_{(\pm,\pm)}^{(n)}Q_{20(\pm)}^{(n)2} 
    & \mp\eta_n f_{(\pm,\pm)}^{(n)}Q_{20(\pm)}^{(n)5}   \\ \hline
  -\frac{1}{\sqrt{2}} f_{(\mp,\pm)}^{(n)}\sigma_{(\pm)}^{(n)1}
    &-\frac{1}{\sqrt{2}} f_{(\mp,\pm)}^{(n)}\sigma_{(\pm)}^{(n)2}  \\
  \ea \right. \non
  \left. 
  \ba{ccc|c}
  \hspace{50pt}
    &\pm\eta_n f_{(\pm,\pm)}^{(n)}Q_{20(\pm)}^{(n)1}  
      &\pm\eta_n f_{(\pm,\pm)}^{(n)}Q_{20(\pm)}^{(n)2}    
        & \frac{1}{\sqrt{2}} f_{(\mp,\pm)}^{(n)}\sigma_{(\pm)}^{(n)1} \\
  \hspace{50pt}
     &\pm\eta_n f_{(\pm,\pm)}^{(n)}Q_{20(\pm)}^{(n)4}  
       &\pm\eta_n f_{(\pm,\pm)}^{(n)}Q_{20(\pm)}^{(n)5}   
        & \frac{1}{\sqrt{2}} f_{(\mp,\pm)}^{(n)}\sigma_{(\pm)}^{(n)2} \\ \hline
     &
       & 
          & \\
     & 
       & \pm\frac{1}{\sqrt{2}} f_{(\pm,\mp)}^{(n)}\zeta_{(\pm)}^{(n)} 
         & \eta_n f_{(\mp,\mp)}^{(n)}U_{20(\pm)}^{(n)1}  \\
     & \mp\frac{1}{\sqrt{2}} f_{(\pm,\mp)}^{(n)}\zeta_{(\pm)}^{(n)} 
       & 
         &\eta_n f_{(\mp,\mp)}^{(n)}U_{20(\pm)}^{(n)2} \\ \hline
      & -\eta_n f_{(\mp,\mp)}^{(n)}U_{20(\pm)}^{(n)1} 
        & -\eta_n f_{(\mp,\mp)}^{(n)}U_{20(\pm)}^{(n)2}  
         &  \\
  \ea\right), \no
 \eea

 \bea
 (\Psi_{(\pm)4})_{jk}=\Psi_{(\pm)4jk}=\frac{1}{\sqrt{6}}\sum_{n=-\infty}^{\infty} \left(
  \ba{cc|}
    &\pm\frac{1}{\sqrt{2}} f_{(\pm,\mp)}^{(n)}\omega_{(\pm)}^{(n)2}  \\
  \mp\frac{1}{\sqrt{2}} f_{(\pm,\mp)}^{(n)}\omega_{(\pm)}^{(n)2}
    &  \\ \hline
  \mp\eta_n f_{(\pm,\pm)}^{(n)}Q_{20(\pm)}^{(n)1}
    &\mp\eta_n f_{(\pm,\pm)}^{(n)}Q_{20(\pm)}^{(n)4}  \\
    &  \\
  \mp\eta_n f_{(\pm,\pm)}^{(n)}Q_{20(\pm)}^{(n)3} 
    & \mp\eta_n f_{(\pm,\pm)}^{(n)}Q_{20(\pm)}^{(n)6}   \\ \hline
  -\frac{1}{\sqrt{2}} f_{(\mp,\pm)}^{(n)}\sigma_{(\pm)}^{(n)3}
    &-\frac{1}{\sqrt{2}} f_{(\mp,\pm)}^{(n)}\sigma_{(\pm)}^{(n)4}  \\
  \ea \right. \non
  \left.
  \ba{ccc|c}
  \pm\eta_n f_{(\pm,\pm)}^{(n)}Q_{20(\pm)}^{(n)1} 
    &\hspace{50pt}
      &\pm\eta_n f_{(\pm,\pm)}^{(n)}Q_{20(\pm)}^{(n)3}    
        & \frac{1}{\sqrt{2}} f_{(\mp,\pm)}^{(n)}\sigma_{(\pm)}^{(n)3} \\
  \pm\eta_n f_{(\pm,\pm)}^{(n)}Q_{20(\pm)}^{(n)4} 
     &  
       &  \pm\eta_n f_{(\pm,\pm)}^{(n)}Q_{20(\pm)}^{(n)6} 
         & \frac{1}{\sqrt{2}} f_{(\mp,\pm)}^{(n)}\sigma_{(\pm)}^{(n)4}   \\ \hline
    &
      & \pm\frac{1}{\sqrt{2}} f_{(\pm,\mp)}^{(n)}\zeta_{(\pm)}^{(n)} 
        & \eta_n f_{(\mp,\mp)}^{(n)}U_{20(\pm)}^{(n)1}  \\
    &
     &
       & \\
  \mp\frac{1}{\sqrt{2}} f_{(\pm,\mp)}^{(n)}\zeta_{(\pm)}^{(n)} 
    &
      &
        &\eta_n f_{(\mp,\mp)}^{(n)}U_{20(\pm)}^{(n)3} \\ \hline
  -\eta_n f_{(\mp,\mp)}^{(n)}U_{20(\pm)}^{(n)1} 
     & 
       & -\eta_n f_{(\mp,\mp)}^{(n)}U_{20(\pm)}^{(n)3}  
         & \\
  \ea \right), \no
 \eea


 \bea
 (\Psi_{(\pm)5})_{jk}=\Psi_{(\pm)5jk}=\frac{1}{\sqrt{6}}\sum_{n=-\infty}^{\infty} \left(
  \ba{cc|}
    &\pm\frac{1}{\sqrt{2}} f_{(\pm,\mp)}^{(n)}\omega_{(\pm)}^{(n)3}  \\
  \mp\frac{1}{\sqrt{2}} f_{(\pm,\mp)}^{(n)}\omega_{(\pm)}^{(n)3} 
    &   \\ \hline
  \mp\eta_n f_{(\pm,\pm)}^{(n)}Q_{20(\pm)}^{(n)2} 
    &\mp\eta_n f_{(\pm,\pm)}^{(n)}Q_{20(\pm)}^{(n)5}  \\
  \mp\eta_n f_{(\pm,\pm)}^{(n)}Q_{20(\pm)}^{(n)3} 
    & \mp\eta_n f_{(\pm,\pm)}^{(n)}Q_{20(\pm)}^{(n)6}  \\
    &  \\ \hline
  -\frac{1}{\sqrt{2}} f_{(\mp,\pm)}^{(n)}\sigma_{(\pm)}^{(n)5}
    &-\frac{1}{\sqrt{2}} f_{(\mp,\pm)}^{(n)}\sigma_{(\pm)}^{(n)6}  \\
  \ea \right.\non
  \left. 
  \ba{ccc|c}
  \pm\eta_n f_{(\pm,\pm)}^{(n)}Q_{20(\pm)}^{(n)2} 
    &\pm\eta_n f_{(\pm,\pm)}^{(n)}Q_{20(\pm)}^{(n)3} 
      &\hspace{50pt} 
        & \frac{1}{\sqrt{2}} f_{(\mp,\pm)}^{(n)}\sigma_{(\pm)}^{(n)5} \\
  \pm\eta_n f_{(\pm,\pm)}^{(n)}Q_{20(\pm)}^{(n)5} 
    &\pm \eta_n f_{(\pm,\pm)}^{(n)}Q_{20(\pm)}^{(n)6}
      &  
        & \frac{1}{\sqrt{2}} f_{(\mp,\pm)}^{(n)}\sigma_{(\pm)}^{(n)6}   \\ \hline
     & \pm\frac{1}{\sqrt{2}} f_{(\pm,\mp)}^{(n)}\zeta_{(\pm)}^{(n)} 
       &  
         & \eta_n f_{(\mp,\mp)}^{(n)}U_{20(\pm)}^{(n)2}  \\
  \mp\frac{1}{\sqrt{2}} f_{(\pm,\mp)}^{(n)}\zeta_{(\pm)}^{(n)} 
     & 
       & 
         & \eta_n f_{(\mp,\mp)}^{(n)}U_{20(\pm)}^{(n)3}  \\
     &
       &
         & \\ \hline
  -\eta_n f_{(\mp,\mp)}^{(n)}U_{20(\pm)}^{(n)2} 
     & -\eta_n f_{(\mp,\mp)}^{(n)}U_{20(\pm)}^{(n)3} 
       &
         & \\
  \ea \right), \no
 \eea


 \bea
 (\Psi_{(\pm)6})_{jk}=\Psi_{(\pm)6jk}=\frac{1}{\sqrt{6}}\sum_{n=-\infty}^{\infty} \left(
  \ba{cc|}
     &\eta_n f_{(\mp,\mp)}^{(n)}\tau_{(\pm)}^{(n)}  \\
  -\eta_n f_{(\mp,\mp)}^{(n)}\tau_{(\pm)}^{(n)}
     &  \\ \hline
  -\frac{1}{\sqrt{2}} f_{(\mp,\pm)}^{(n)}\sigma_{(\pm)}^{(n)1} 
     &-\frac{1}{\sqrt{2}} f_{(\mp,\pm)}^{(n)}\sigma_{(\pm)}^{(n)2} \\
  -\frac{1}{\sqrt{2}} f_{(\mp,\pm)}^{(n)}\sigma_{(\pm)}^{(n)3} 
     & -\frac{1}{\sqrt{2}} f_{(\mp,\pm)}^{(n)}\sigma_{(\pm)}^{(n)4}\\
  -\frac{1}{\sqrt{2}} f_{(\mp,\pm)}^{(n)}\sigma_{(\pm)}^{(n)4}
     & -\frac{1}{\sqrt{2}} f_{(\mp,\pm)}^{(n)}\sigma_{(\pm)}^{(n)6}  \\ \hline
     & \\
  \ea \right.\non
  \left. 
  \ba{ccc|c}
  \frac{1}{\sqrt{2}} f_{(\mp,\pm)}^{(n)}\sigma_{(\pm)}^{(n)1} 
    &\frac{1}{\sqrt{2}} f_{(\mp,\pm)}^{(n)}\sigma_{(\pm)}^{(n)3}
      &\frac{1}{\sqrt{2}} f_{(\mp,\pm)}^{(n)}\sigma_{(\pm)}^{(n)5}  
        & \hspace{50pt}\\
  \frac{1}{\sqrt{2}} f_{(\mp,\pm)}^{(n)}\sigma_{(\pm)}^{(n)2}
    & \frac{1}{\sqrt{2}} f_{(\mp,\pm)}^{(n)}\sigma_{(\pm)}^{(n)4}
      &  \frac{1}{\sqrt{2}} f_{(\mp,\pm)}^{(n)}\sigma_{(\pm)}^{(n)6} 
        &  \\\hline
     & \eta_n f_{(\mp,\mp)}^{(n)}U_{20(\pm)}^{(n)1}
       & \eta_n f_{(\mp,\mp)}^{(n)}U_{20(\pm)}^{(n)2}
         &   \\
  -\eta_n f_{(\mp,\mp)}^{(n)}U_{20(\pm)}^{(n)1} 
      & 
        & \eta_n f_{(\mp,\mp)}^{(n)}U_{20(\pm)}^{(n)3} 
          &  \\
  -\eta_n f_{(\mp,\mp)}^{(n)}U_{20(\pm)}^{(n)2} 
       & -\eta_n f_{(\mp,\pm)}^{(n)}U_{20(\pm)}^{(n)3}
         &  
           & \\ \hline
    &
      &
        & \\
  \ea \right). \no
 \eea

The diagonalized mass terms are derived as  
 \bea
 \lag_4 \supset
  -\sum_{n=-\infty}^{\infty}\left[
            \sum_{i=1}^{3}\frac{n+\alpha}{R} \overline{\Psi}^{(n)i}_{(\pm)}\Psi^{(n)i}_{(\pm)}
            +\sum_{i=4}^{6}\frac{n+\alpha+1/2}{R} \overline{\Psi}^{(n)i}_{(\pm)}\Psi^{(n)i}_{(\pm)}\right]\non
  -\sum_{n=0}^{\infty}\left[\sum_{i=7}^{10}\frac{n}{R} \overline{\Psi}^{(n)i}_{(\pm)}\Psi^{i(n)}_{(\pm)}
      +\sum_{i=11}^{14}\frac{n+1/2}{R} \overline{\Psi}^{(n)i}_{(\pm)}\Psi^{(n)i}_{(\pm)}\right] 
 \eea
and the corresponding mass eigenstates are found
 \bea
 &&\Psi^{(n)\{1,2,3\}}_{(\pm)}=\eta_n\left(Q^{(n)\{4,5,6\}}_{20(\pm)}
                                                                     -U^{(n)\{1,2,3\}}_{20(\pm)}\right), \quad 
   \Psi^{(n)\{4,5,6\}}_{(\pm)}=\frac{1}{\sqrt{2}}\left(\omega^{(n)\{1,2,3\}}_{(\pm)}
                                                                     -\sigma^{(n)\{1,2,3\}}_{(\pm)}\right)\non
 &&\Psi^{(n)7}_{(\pm)}=\tau^{(n)}_{(\pm)}, \quad 
   \Psi^{(n)\{8,9,10\}}_{(\pm)}=Q^{(n)\{1,2,3\}}_{20(\pm)}, \quad
 \Psi^{(n)\{11,12,13\}}_{(\pm)}=\sigma^{(n)\{4,5,6\}}_{(\pm)}, \quad 
   \Psi^{(n)14}_{(\pm)}=\zeta^{(n)}_{(\pm)}. 
 \eea

\subsubsection{56: the third rank symmetric tensor}
The ${\bf 56}$ representation is the third rank symmetric tensor of $SU(6)$. 
The components after the decomposition into $SU(3)_C \times SU(2)_L \times U(1)_Y \times U(1)_X$ 
and the corresponding parity and reflection are summarized in Table \ref{table:bulk parity 56}.

 \begin{table}[h]
  \centering 
   \begin{tabular}{|c|c||c|c|} \hline
   $(+,+)$
    &
     &$(+,-)$
      &\\ \hline
  $(3, 2)$
     &$Q^{(-n)}_{56(\pm)}=\mp Q^{(n)}_{56(\pm)}$
       &$(1,1)$
         &$\chi^{(-n-1)}_{(\pm)}=\mp\chi^{(n)}_{(\pm)}$\\ \hline
    &
      &$(6, 1)$
        &$\rho^{(-n-1)}_{(\pm)}=\mp\rho^{(n)}_{(\pm)}$\\ \hline
    &
      &$(1, 3)$
        &$\tau^{(-n-1)}_{(\pm)}=\mp\tau^{(n)}_{(\pm)}$\\ \hline \hline
  $(-,-)$
    &  
     &$(-,+)$
       &\\ \hline
  $(3, 3)$
    &$\omega^{(-n)}_{(\pm)}=\pm\omega^{(n)}_{(\pm)}$
     &$(6, 2)$
      &$\theta^{(-n-1)}_{(\pm)}=\pm\theta^{(n)}_{(\pm)}$\\ \hline
  $(10, 1)$
    &$\zeta^{(-n)}_{(\pm)}=\pm\zeta^{(n)}_{(\pm)}$
      &$(1,4)$
       &$\phi^{(-n-1)}_{(\pm)}=\pm\phi^{(n)}_{(\pm)}$\\ \hline
  $(3, 1)$
     &$D^{(-n)}_{56(\pm)}=\pm D^{(n)}_{56(\pm)}$
       &$(1,2)$
         &$\nu^{(-n-1)}_{(\pm)}=\pm\nu^{(n)}_{(\pm)}$\\ \hline
  \end{tabular} \non
  \caption{Parity and reflection for  inner component of ${\bf 56}$.}
  \label{table:bulk parity 56}
 \end{table}

KK expansion and the diagonalization of mass matrix can be done 
 similarly as in B.2.3. 

 \bea
 (\Psi_{(\pm)1})_{jk}=\Psi_{(\pm)1jk}=\frac{1}{\sqrt{6}}\sum_{n=-\infty}^{\infty} \left( 
  \ba{cc|}
  \pm\sqrt{6}\frac{1}{\sqrt{2}}f_{(\pm,\mp)}^{(n)}\phi_{(\pm)}^{(n)1}
      & \pm\sqrt{2}\frac{1}{\sqrt{2}}f_{(\pm,\mp)}^{(n)}\phi_{(\pm)}^{(n)2} \\
  \pm\sqrt{2}\frac{1}{\sqrt{2}}f_{(\pm,\mp)}^{(n)}\phi_{(\pm)}^{(n)2} 
      & \pm\sqrt{2}\frac{1}{\sqrt{2}}f_{(\pm,\mp)}^{(n)}\phi_{(\pm)}^{(n)3}   \\ \hline
  \pm\sqrt{2}\eta_n f_{(\pm,\pm)}^{(n)}\omega_{(\pm)}^{(n)1} 
      & \pm\eta_n f_{(\pm,\pm)}^{(n)}\omega_{(\pm)}^{(n)2} \\
  \pm\sqrt{2}\eta_n f_{(\pm,\pm)}^{(n)}\omega_{(\pm)}^{(n)3}
      & \pm\eta_n f_{(\pm,\pm)}^{(n)}\omega_{(\pm)}^{(n)4}\\
  \pm\sqrt{2}\eta_n f_{(\pm,\pm)}^{(n)}\omega_{(\pm)}^{(n)5} 
      & \pm\eta_n f_{(\pm,\pm)}^{(n)}\omega_{(\pm)}^{(n)6}  \\ \hline
  \sqrt{2}\frac{1}{\sqrt{2}} f_{(\mp,\pm)}^{(n)}\tau_{(\pm)}^{(n)1}  
      &  \frac{1}{\sqrt{2}} f_{(\mp,\pm)}^{(n)}\tau_{(\pm)}^{(n)2} 
  \ea \right. \non
  \left. 
  \ba{ccc|c}
  \pm\sqrt{2}\eta_n f_{(\pm,\pm)}^{(n)}\omega_{(\pm)}^{(n)1} 
     & \pm\sqrt{2}\eta_n f_{(\pm,\pm)}^{(n)}\omega_{(\pm)}^{(n)3}
       & \pm\sqrt{2}\eta_n f_{(\pm,\pm)}^{(n)}\omega_{(\pm)}^{(n)5} 
          &\sqrt{2}\frac{1}{\sqrt{2}} f_{(\mp,\pm)}^{(n)}\tau_{(\pm)}^{(n)1} \\
  \pm\eta_n f_{(\pm,\pm)}^{(n)}\omega_{(\pm)}^{(n)2}
     & \pm\eta_n f_{(\pm,\pm)}^{(n)}\omega_{(\pm)}^{(n)4}
        & \pm\eta_n f_{(\pm,\pm)}^{(n)}\omega_{(\pm)}^{(n)6} 
           & \frac{1}{\sqrt{2}} f_{(\mp,\pm)}^{(n)}\tau_{(\pm)}^{(n)2} \\ \hline
  \pm\sqrt{2}\frac{1}{\sqrt{2}} f_{(\pm,\mp)}^{(n)}\theta_{(\pm)}^{(n)1} 
     & \pm\sqrt{2}\frac{1}{\sqrt{2}} f_{(\pm,\mp)}^{(n)}\theta_{(\pm)}^{(n)2} 
        &\pm\frac{1}{\sqrt{2}} f_{(\pm,\mp)}^{(n)}\theta_{(\pm)}^{(n)4} 
          & \eta_n f_{(\mp,\mp)}^{(n)}Q_{56(\pm)}^{\ast(n)1} \\
  \pm\frac{1}{\sqrt{2}} f_{(\pm,\mp)}^{(n)}\theta_{(\pm)}^{(n)2} 
     &\pm\frac{1}{\sqrt{2}} f_{(\pm,\mp)}^{(n)}\theta_{(\pm)}^{(n)3}
        & \pm\sqrt{2}\frac{1}{\sqrt{2}} f_{(\pm,\mp)}^{(n)}\theta_{(\pm)}^{(n)5} 
          & \eta_n f_{(\mp,\mp)}^{(n)}Q_{56(\pm)}^{\ast(n)3} \\
 \pm\frac{1}{\sqrt{2}} f_{(\pm,\mp)}^{(n)}\theta_{(\pm)}^{(n)4} 
     & \pm\frac{1}{\sqrt{2}} f_{(\pm,\mp)}^{(n)}\theta_{(\pm)}^{(n)5}
        & \pm\sqrt{2}\frac{1}{\sqrt{2}} f_{(\pm,\mp)}^{(n)}\theta_{(\pm)}^{(n)6} 
          & \eta_n f_{(\mp,\mp)}^{(n)}Q_{56(\pm)}^{\ast(n)5} \\ \hline
  \eta_n f_{(\mp,\mp)}^{(n)}Q_{56(\pm)}^{\ast(n)1} 
     & \eta_n f_{(\mp,\mp)}^{(n)}Q_{56(\pm)}^{\ast(n)3}  
        & \eta_n f_{(\mp,\mp)}^{(n)}Q_{56(\pm)}^{\ast(n)5}  
           & \pm\sqrt{2} \frac{1}{\sqrt{2}} f_{(\pm,\mp)}^{(n)}\nu_{(\pm)}^{(n)1} \\
  \ea \right), \no
 \eea


 \bea
 (\Psi_{(\pm)2})_{jk}=\Psi_{(\pm)2jk}=\frac{1}{\sqrt{6}}\sum_{n=-\infty}^{\infty} \left(
   \ba{cc|}
   \pm\sqrt{2}\frac{1}{\sqrt{2}}f_{(\pm,\mp)}^{(n)}\phi_{(\pm)}^{(n)2} 
       & \pm\sqrt{2}\frac{1}{\sqrt{2}}f_{(\pm,\mp)}^{(n)}\phi_{(\pm)}^{(n)3} \\
  \pm\sqrt{2}\frac{1}{\sqrt{2}}f_{(\pm,\mp)}^{(n)}\phi_{(\pm)}^{(n)3}  
       & \pm\sqrt{6}\frac{1}{\sqrt{2}}f_{(\pm,\mp)}^{(n)}\phi_{(\pm)}^{(n)4}   \\ \hline
  \pm\eta_n f_{(\pm,\pm)}^{(n)}\omega_{(\pm)}^{(n)2} 
       & \pm\sqrt{2}\eta_n f_{(\pm,\pm)}^{(n)}\omega_{(\pm)}^{(n)7} \\
  \pm\eta_n f_{(\pm,\pm)}^{(n)}\omega_{(\pm)}^{(n)4}
       & \pm\sqrt{2}\eta_n f_{(\pm,\pm)}^{(n)}\omega_{(\pm)}^{(n)8}\\
  \pm\eta_n f_{(\pm,\pm)}^{(n)}\omega_{(\pm)}^{(n)6} 
       & \pm\sqrt{2}\eta_n f_{(\pm,\pm)}^{(n)}\omega_{(\pm)}^{(n)9}  \\ \hline
  \frac{1}{\sqrt{2}} f_{(\mp,\pm)}^{(n)}\tau_{(\pm)}^{(n)2}  
       & \sqrt{2}\frac{1}{\sqrt{2}} f_{(\mp,\pm)}^{(n)}\tau_{(\pm)}^{(n)3}  \\
  \ea \right. \non
  \left. 
  \ba{ccc|c}
  \pm\eta_n f_{(\pm,\pm)}^{(n)}\omega_{(\pm)}^{(n)2} 
       & \pm\eta_n f_{(\pm,\pm)}^{(n)}\omega_{(\pm)}^{(n)4}
            & \pm\eta_n f_{(\pm,\pm)}^{(n)}\omega_{(\pm)}^{(n)6}  
                &\frac{1}{\sqrt{2}} f_{(\mp,\pm)}^{(n)}\tau_{(\pm)}^{(n)2} \\
  \pm\eta_n f_{(\pm,\pm)}^{(n)}\omega_{(\pm)}^{(n)7} 
       & \pm\eta_n f_{(\pm,\pm)}^{(n)}\omega_{(\pm)}^{(n)8}
            & \pm\eta_n f_{(\pm,\pm)}^{(n)}\omega_{(\pm)}^{(n)9}  
                & \sqrt{2}\frac{1}{\sqrt{2}} f_{(\mp,\pm)}^{(n)}\tau_{(\pm)}^{(n)3} \\ \hline
  \pm\sqrt{2}\frac{1}{\sqrt{2}} f_{(\pm,\mp)}^{(n)}\theta_{(\pm)}^{(n)7} 
       & \pm\sqrt{2}\frac{1}{\sqrt{2}} f_{(\pm,\mp)}^{(n)}\theta_{(\pm)}^{(n)8} 
           &\pm\frac{1}{\sqrt{2}} f_{(\pm,\mp)}^{(n)}\theta_{(\pm)}^{(n)10} 
                & \eta_n f_{(\mp,\mp)}^{(n)}Q_{56(\pm)}^{\ast(n)2} \\
  \pm\frac{1}{\sqrt{2}} f_{(\pm,\mp)}^{(n)}\theta_{(\pm)}^{(n)8} 
       &\pm\frac{1}{\sqrt{2}} f_{(\pm,\mp)}^{(n)}\theta_{(\pm)}^{(n)9}
           & \pm\sqrt{2}\frac{1}{\sqrt{2}} f_{(\pm,\mp)}^{(n)}\theta_{(\pm)}^{(n)11} 
              & \eta_n f_{(\mp,\mp)}^{(n)}Q_{56(\pm)}^{\ast(n)4} \\
  \pm\frac{1}{\sqrt{2}} f_{(\pm,\mp)}^{(n)}\theta_{(\pm)}^{(n)10} 
       & \pm\frac{1}{\sqrt{2}} f_{(\pm,\mp)}^{(n)}\theta_{(\pm)}^{(n)11}
           & \pm\sqrt{2}\frac{1}{\sqrt{2}} f_{(\pm,\mp)}^{(n)}\theta_{(\pm)}^{(n)12} 
             & \eta_n f_{(\mp,\mp)}^{(n)}Q_{56(\pm)}^{\ast(n)6} \\ \hline
  \eta_n f_{(\mp,\mp)}^{(n)}Q_{56(\pm)}^{\ast(n)2} 
       & \eta_n f_{(\mp,\mp)}^{(n)}Q_{56(\pm)}^{\ast(n)4} 
          & \eta_n f_{(\mp,\mp)}^{(n)}Q_{56(\pm)}^{\ast(n)6}  
              & \pm\sqrt{2} \frac{1}{\sqrt{2}} f_{(\pm,\mp)}^{(n)}\nu_{(\pm)}^{(n)2} \\
  \ea \right),\no
 \eea


 \bea
 (\Psi_{(\pm)3})_{jk}=\Psi_{(\pm)3jk}=\frac{1}{\sqrt{6}}\sum_{n=-\infty}^{\infty} \left( 
  \ba{cc|}
  \pm\sqrt{2}\eta_n f_{(\pm,\pm)}^{(n)}\omega_{(\pm)}^{(n)1}
      &\pm\eta_n f_{(\pm,\pm)}^{(n)}\omega_{(\pm)}^{(n)2}  \\
  \pm\eta_n f_{(\pm,\pm)}^{(n)}\omega_{(\pm)}^{(n)2} 
      & \pm\sqrt{2}\eta_n f_{(\pm,\pm)}^{(n)}\omega_{(\pm)}^{(n)7}    \\ \hline
  \pm\sqrt{2}\frac{1}{\sqrt{2}} f_{(\pm,\mp)}^{(n)}\theta_{(\pm)}^{(n)1} 
      &\pm\sqrt{2}\frac{1}{\sqrt{2}} f_{(\pm,\mp)}^{(n)}\theta_{(\pm)}^{(n)7}  \\
  \pm\frac{1}{\sqrt{2}} f_{(\pm,\mp)}^{(n)}\theta_{(\pm)}^{(n)2} 
      & \pm\frac{1}{\sqrt{2}} f_{(\pm,\mp)}^{(n)}\theta_{(\pm)}^{(n)8}  \\
  \pm\frac{1}{\sqrt{2}} f_{(\pm,\mp)}^{(n)}\theta_{(\pm)}^{(n)4} 
      & \pm\frac{1}{\sqrt{2}} f_{(\pm,\mp)}^{(n)}\theta_{(\pm)}^{(n)10}   \\ \hline
  \eta_n f_{(\mp,\mp)}^{(n)}Q_{56(\pm)}^{\ast(n)1} 
      &\eta_n f_{(\mp,\mp)}^{(n)}Q_{56(\pm)}^{\ast(n)2}  \\
  \ea \right. \non
  \left. 
  \ba{ccc|c}
  \pm\sqrt{2}\frac{1}{\sqrt{2}} f_{(\pm,\mp)}^{(n)}\theta_{(\pm)}^{(n)1} 
     &\pm\frac{1}{\sqrt{2}} f_{(\pm,\mp)}^{(n)}\theta_{(\pm)}^{(n)2}  
       &\pm\frac{1}{\sqrt{2}} f_{(\pm,\mp)}^{(n)}\theta_{(\pm)}^{(n)4}    
         & \eta_n f_{(\mp,\mp)}^{(n)}Q_{56(\pm)}^{\ast(n)1} \\
  \pm\sqrt{2}\frac{1}{\sqrt{2}} f_{(\pm,\mp)}^{(n)}\theta_{(\pm)}^{(n)7} 
     & \pm\frac{1}{\sqrt{2}} f_{(\pm,\mp)}^{(n)}\theta_{(\pm)}^{(n)8} 
        &  \pm\frac{1}{\sqrt{2}} f_{(\pm,\mp)}^{(n)}\theta_{(\pm)}^{(n)10}  
           & \eta_n f_{(\mp,\mp)}^{(n)}Q_{56(\pm)}^{\ast(n)2}   \\ \hline
  \pm\sqrt{6}\eta_n f_{(\pm,\pm)}^{(n)}\zeta_{(\pm)}^{(n)1}
     & \pm\sqrt{2}\eta_n f_{(\pm,\pm)}^{(n)}\zeta_{(\pm)}^{(n)2} 
        & \pm\sqrt{2}\eta_n f_{(\pm,\pm)}^{(n)}\zeta_{(\pm)}^{(n)4} 
           & \sqrt{2}\frac{1}{\sqrt{2}} f_{(\mp,\pm)}^{(n)}\rho_{(\pm)}^{(n)1}  \\
  \pm\sqrt{2}\eta_n f_{(\pm,\pm)}^{(n)}\zeta_{(\pm)}^{(n)2} 
     & \pm\sqrt{2}\eta_n f_{(\pm,\pm)}^{(n)}\zeta_{(\pm)}^{(n)3} 
        & \pm\eta_n f_{(\pm,\pm)}^{(n)}\zeta_{(\pm)}^{(n)5} 
           & \frac{1}{\sqrt{2}} f_{(\mp,\pm)}^{(n)}\rho_{(\pm)}^{(n)2}  \\
  \pm\sqrt{2}\eta_n f_{(\pm,\pm)}^{(n)}\zeta_{(\pm)}^{(n)4} 
     & \pm\eta_n f_{(\pm,\pm)}^{(n)}\zeta_{(\pm)}^{(n)5} 
        & \pm\sqrt{2}\eta_n f_{(\pm,\pm)}^{(n)}\zeta_{(\pm)}^{(n)6} 
           & \frac{1}{\sqrt{2}} f_{(\mp,\pm)}^{(n)}\rho_{(\pm)}^{(n)4} \\ \hline
  \sqrt{2}\frac{1}{\sqrt{2}} f_{(\mp,\pm)}^{(n)}\rho_{(\pm)}^{(n)1} 
     & \frac{1}{\sqrt{2}} f_{(\mp,\pm)}^{(n)}\rho_{(\pm)}^{(n)2} 
        & \frac{1}{\sqrt{2}} f_{(\mp,\pm)}^{(n)}\rho_{(\pm)}^{(n)4}  
           & \pm\eta_n\sqrt{2} f_{(\pm,\pm)}^{(n)}D_{\ast(\pm)}^{56(n)1} \\
  \ea \right),\no
 \eea


 \bea
 (\Psi_{(\pm)4})_{jk}=\Psi_{(\pm)4jk}=\frac{1}{\sqrt{6}}\sum_{n=-\infty}^{\infty} \left(
  \ba{cc|}
  \pm\sqrt{2}\eta_n f_{(\pm,\pm)}^{(n)}\omega_{(\pm)}^{(n)3}
      &\pm\eta_n f_{(\pm,\pm)}^{(n)}\omega_{(\pm)}^{(n)4}  \\
  \pm\eta_n f_{(\pm,\pm)}^{(n)}\omega_{(\pm)}^{(n)4} 
      & \pm\sqrt{2}\eta_n f_{(\pm,\pm)}^{(n)}\omega_{(\pm)}^{(n)8}    \\ \hline
  \pm\frac{1}{\sqrt{2}} f_{(\pm,\mp)}^{(n)}\theta_{(\pm)}^{(n)2} 
      &\pm\frac{1}{\sqrt{2}} f_{(\pm,\mp)}^{(n)}\theta_{(\pm)}^{(n)8}  \\
  \pm\sqrt{2}\frac{1}{\sqrt{2}} f_{(\pm,\mp)}^{(n)}\theta_{(\pm)}^{(n)3} 
      & \pm\sqrt{2}\frac{1}{\sqrt{2}} f_{(\pm,\mp)}^{(n)}\theta_{(\pm)}^{(n)9}  \\
  \pm\frac{1}{\sqrt{2}} f_{(\pm,\mp)}^{(n)}\theta_{(\pm)}^{(n)5} 
      & \pm\frac{1}{\sqrt{2}} f_{(\pm,\mp)}^{(n)}\theta_{(\pm)}^{(n)11}   \\ \hline
  \eta_n f_{(\mp,\mp)}^{(n)}Q_{56(\pm)}^{\ast(n)3} 
      &\eta_n f_{(\mp,\mp)}^{(n)}Q_{56(\pm)}^{\ast(n)4}  \\
  \ea \right. \non
  \left. 
  \ba{ccc|c}
   \pm\frac{1}{\sqrt{2}} f_{(\pm,\mp)}^{(n)}\theta_{(\pm)}^{(n)2} 
      &\pm\sqrt{2}\frac{1}{\sqrt{2}} f_{(\pm,\mp)}^{(n)}\theta_{(\pm)}^{(n)3}  
         &\pm\frac{1}{\sqrt{2}} f_{(\pm,\mp)}^{(n)}\theta_{(\pm)}^{(n)5}    
            & \eta_n f_{(\mp,\mp)}^{(n)}Q_{56(\pm)}^{\ast(n)3} \\
  \pm\frac{1}{\sqrt{2}} f_{(\pm,\mp)}^{(n)}\theta_{(\pm)}^{(n)8} 
      & \pm\sqrt{2}\frac{1}{\sqrt{2}} f_{(\pm,\mp)}^{(n)}\theta_{(\pm)}^{(n)9}
         &  \pm\frac{1}{\sqrt{2}} f_{(\pm,\mp)}^{(n)}\theta_{(\pm)}^{(n)11}  
           & \eta_n f_{(\mp,\mp)}^{(n)}Q_{56(\pm)}^{\ast(n)4}   \\ \hline
  \pm\sqrt{2}\eta_n f_{(\pm,\pm)}^{(n)}\zeta_{(\pm)}^{(n)2}
     & \pm\sqrt{2}\eta_n f_{(\pm,\pm)}^{(n)}\zeta_{(\pm)}^{(n)3} 
        & \pm\eta_n f_{(\pm,\pm)}^{(n)}\zeta_{(\pm)}^{(n)5} 
           & \frac{1}{\sqrt{2}} f_{(\mp,\pm)}^{(n)}\rho_{(\pm)}^{(n)2}  \\
  \pm\sqrt{2}\eta_n f_{(\pm,\pm)}^{(n)}\zeta_{(\pm)}^{(n)3}
      & \pm\sqrt{6}\eta_n f_{(\pm,\pm)}^{(n)}\zeta_{(\pm)}^{(n)7} 
         & \pm\sqrt{2}\eta_n f_{(\pm,\pm)}^{(n)}\zeta_{(\pm)}^{(n)8} 
            & \sqrt{2}\frac{1}{\sqrt{2}} f_{(\mp,\pm)}^{(n)}\rho_{(\pm)}^{(n)3}  \\
  \pm\eta_n f_{(\pm,\pm)}^{(n)}\zeta_{(\pm)}^{(n)5} 
      & \pm\sqrt{2}\eta_n f_{(\pm,\pm)}^{(n)}\zeta_{(\pm)}^{(n)8} 
         & \pm\sqrt{2}\eta_n f_{(\pm,\pm)}^{(n)}\zeta_{(\pm)}^{(n)9} 
            & \frac{1}{\sqrt{2}} f_{(\mp,\pm)}^{(n)}\rho_{(\pm)}^{(n)5} \\ \hline
  \frac{1}{\sqrt{2}} f_{(\mp,\pm)}^{(n)}\rho_{(\pm)}^{(n)2} 
      & \sqrt{2}\frac{1}{\sqrt{2}} f_{(\mp,\pm)}^{(n)}\rho_{(\pm)}^{(n)3} 
         & \frac{1}{\sqrt{2}} f_{(\mp,\pm)}^{(n)}\rho_{(\pm)}^{(n)5}  
           & \pm\sqrt{2}\eta_n f_{(\pm,\pm)}^{(n)}D_{\ast(\pm)}^{56(n)2} \\
  \ea \right), \no
 \eea


 \bea
 (\Psi_{(\pm)5})_{jk}=\Psi_{(\pm)5jk}=\frac{1}{\sqrt{6}}\sum_{n=-\infty}^{\infty} \left(
  \ba{cc|}
  \pm\sqrt{2}\eta_n f_{(\pm,\pm)}^{(n)}\omega_{(\pm)}^{(n)5}
     &\pm\eta_n f_{(\pm,\pm)}^{(n)}\omega_{(\pm)}^{(n)6}  \\
  \pm\eta_n f_{(\pm,\pm)}^{(n)}\omega_{(\pm)}^{(n)6} 
     & \pm\sqrt{2}\eta_n f_{(\pm,\pm)}^{(n)}\omega_{(\pm)}^{(n)9}    \\ \hline
  \pm\frac{1}{\sqrt{2}} f_{(\pm,\mp)}^{(n)}\theta_{(\pm)}^{(n)4} 
     &\pm\frac{1}{\sqrt{2}} f_{(\pm,\mp)}^{(n)}\theta_{(\pm)}^{(n)10}  \\
  \pm\frac{1}{\sqrt{2}} f_{(\pm,\mp)}^{(n)}\theta_{(\pm)}^{(n)5} 
     & \pm\frac{1}{\sqrt{2}} f_{(\pm,\mp)}^{(n)}\theta_{(\pm)}^{(n)11}  \\
  \pm\sqrt{2}\frac{1}{\sqrt{2}} f_{(\pm,\mp)}^{(n)}\theta_{(\pm)}^{(n)6} 
     & \pm\sqrt{2}\frac{1}{\sqrt{2}} f_{(\pm,\mp)}^{(n)}\theta_{(\pm)}^{(n)12}   \\ \hline
  \eta_n f_{(\mp,\mp)}^{(n)}Q_{56(\pm)}^{\ast(n)5} 
     &\eta_n f_{(\mp,\mp)}^{(n)}Q_{56(\pm)}^{\ast(n)6}  \\
  \ea \right.\non
  \left. 
  \ba{ccc|c}
  \pm\frac{1}{\sqrt{2}} f_{(\pm,\mp)}^{(n)}\theta_{(\pm)}^{(n)4} 
     &\pm\frac{1}{\sqrt{2}} f_{(\pm,\mp)}^{(n)}\theta_{(\pm)}^{(n)5}  
        &\pm\sqrt{2}\frac{1}{\sqrt{2}} f_{(\pm,\mp)}^{(n)}\theta_{(\pm)}^{(n)6}    
           & \eta_n f_{(\mp,\mp)}^{(n)}Q_{56(\pm)}^{\ast(n)5} \\
  \pm\frac{1}{\sqrt{2}} f_{(\pm,\mp)}^{(n)}\theta_{(\pm)}^{(n)10}
     & \pm\frac{1}{\sqrt{2}} f_{(\pm,\mp)}^{(n)}\theta_{(\pm)}^{(n)11} 
       &  \pm\sqrt{2}\frac{1}{\sqrt{2}} f_{(\pm,\mp)}^{(n)}\theta_{(\pm)}^{(n)12}  
         & \eta_n f_{(\mp,\mp)}^{(n)}Q_{56(\pm)}^{\ast(n)6}   \\ \hline
  \pm\sqrt{2}\eta_n f_{(\pm,\pm)}^{(n)}\zeta_{(\pm)}^{(n)4}
     & \pm\eta_n f_{(\pm,\pm)}^{(n)}\zeta_{(\pm)}^{(n)5} 
       & \pm\sqrt{2}\eta_n f_{(\pm,\pm)}^{(n)}\zeta_{(\pm)}^{(n)6} 
         & \frac{1}{\sqrt{2}} f_{(\mp,\pm)}^{(n)}\rho_{(\pm)}^{(n)4}  \\
  \pm\eta_n f_{(\pm,\pm)}^{(n)}\zeta_{(\pm)}^{(n)5}
     & \pm\sqrt{2}\eta_n f_{(\pm,\pm)}^{(n)}\zeta_{(\pm)}^{(n)8} 
       & \pm\sqrt{2}\eta_n f_{(\pm,\pm)}^{(n)}\zeta_{(\pm)}^{(n)9} 
         & \frac{1}{\sqrt{2}} f_{(\mp,\pm)}^{(n)}\rho_{(\pm)}^{(n)5}  \\
  \pm\sqrt{2}\eta_n f_{(\pm,\pm)}^{(n)}\zeta_{(\pm)}^{(n)6} 
     & \pm\sqrt{2}\eta_n f_{(\pm,\pm)}^{(n)}\zeta_{(\pm)}^{(n)9} 
       & \pm\sqrt{6}\eta_n f_{(\pm,\pm)}^{(n)}\zeta_{(\pm)}^{(n)10} ]
         & \sqrt{2}\frac{1}{\sqrt{2}} f_{(\mp,\pm)}^{(n)}\rho_{(\pm)}^{(n)6} \\ \hline
  \frac{1}{\sqrt{2}} f_{(\mp,\pm)}^{(n)}\rho_{(\pm)}^{(n)4} 
     & \frac{1}{\sqrt{2}} f_{(\mp,\pm)}^{(n)}\rho_{(\pm)}^{(n)5} 
       & \sqrt{2}\frac{1}{\sqrt{2}} f_{(\mp,\pm)}^{(n)}\rho_{(\pm)}^{(n)6}  
         & \pm\sqrt{2}\eta_n f_{(\pm,\pm)}^{(n)}D_{\ast(\pm)}^{56(n)3} \\
  \ea \right), \no
 \eea


 \bea
 (\Psi_{(\pm)6})_{jk}=\Psi_{(\pm)6jk}=\frac{1}{\sqrt{6}}\sum_{n=-\infty}^{\infty} \left(
  \ba{cc|}
  \sqrt{2}\frac{1}{\sqrt{2}}f_{(\mp,\pm)}^{(n)}\tau_{(\pm)}^{(n)1}
     &\frac{1}{\sqrt{2}}f_{(\mp,\pm)}^{(n)}\tau_{(\pm)}^{(n)2}  \\
  \frac{1}{\sqrt{2}}f_{(\mp,\pm)}^{(n)}\tau_{(\pm)}^{(n)2}
     & \sqrt{2}\frac{1}{\sqrt{2}}f_{(\mp,\pm)}^{(n)}\tau_{(\pm)}^{(n)3}   \\ \hline
  \eta_n f_{(\mp,\mp)}^{(n)}Q_{56(\pm)}^{\ast(n)1} 
     &\eta_n f_{(\mp,\mp)}^{(n)}Q_{56(\pm)}^{\ast(n)2} \\
  \eta_n f_{(\mp,\mp)}^{(n)}Q_{56(\pm)}^{\ast(n)3} 
     & \eta_n f_{(\mp,\mp)}^{(n)}Q_{56(\pm)}^{\ast(n)4}\\
  \eta_n f_{(\mp,\mp)}^{(n)}Q_{56(\pm)}^{\ast(n)4}
     & \eta_n f_{(\mp,\mp)}^{(n)}Q_{56(\pm)}^{\ast(n)6}  \\ \hline
  \pm\sqrt{2}\frac{1}{\sqrt{2}} f_{(\pm,\mp)}^{(n)}\nu_{(\pm)}^{(n)1} 
     &\pm\sqrt{2}\frac{1}{\sqrt{2}} f_{(\pm,\mp)}^{(n)}\nu_{(\pm)}^{(n)2} \\
  \ea \right.\non
  \left. 
  \ba{ccc|c}
  \eta_n f_{(\mp,\mp)}^{(n)}Q_{56(\pm)}^{\ast(n)1} 
     &\eta_n f_{(\mp,\mp)}^{(n)}Q_{56(\pm)}^{\ast(n)3}
       &\eta_n f_{(\mp,\mp)}^{(n)}Q_{56(\pm)}^{\ast(n)5}  
         & \pm\sqrt{2}\frac{1}{\sqrt{2}} f_{(\pm,\mp)}^{(n)}\nu_{(\pm)}^{(n)1}\\
  \eta_n f_{(\mp,\mp)}^{(n)}Q_{56(\pm)}^{\ast(n)2}
     & \eta_n f_{(\mp,\mp)}^{(n)}Q_{56(\pm)}^{\ast(n)4}
       &  \eta_n f_{(\mp,\mp)}^{(n)}Q_{56(\pm)}^{\ast(n)6} 
        & \pm\sqrt{2}\frac{1}{\sqrt{2}} f_{(\pm,\mp)}^{(n)}\nu_{(\pm)}^{(n)2}  \\ \hline
  \sqrt{2}\frac{1}{\sqrt{2}} f_{(\mp,\pm)}^{(n)}\rho_{(\pm)}^{(n)1}
     & \frac{1}{\sqrt{2}} f_{(\mp,\pm)}^{(n)}\rho_{(\pm)}^{(n)2}
       & \frac{1}{\sqrt{2}} f_{(\mp,\pm)}^{(n)}\rho_{(\pm)}^{(n)4}
         & \pm\sqrt{2}\eta_n f_{(\pm,\pm)}^{(n)}D_{56(\pm)}^{\ast(n)1}  \\
  \frac{1}{\sqrt{2}} f_{(\mp,\pm)}^{(n)}\rho_{(\pm)}^{(n)2} 
     & \sqrt{2}\frac{1}{\sqrt{2}} f_{(\mp,\pm)}^{(n)}\rho_{(\pm)}^{(n)3}
       & \frac{1}{\sqrt{2}} f_{(\mp,\pm)}^{(n)}\rho_{(\pm)}^{(n)5} 
         & \pm\sqrt{2}\eta_n f_{(\pm,\pm)}^{(n)}D_{56(\pm)}^{\ast(n)2}   \\
  \frac{1}{\sqrt{2}} f_{(\mp,\pm)}^{(n)}\rho_{(\pm)}^{(n)4} 
     & \frac{1}{\sqrt{2}} f_{(\mp,\pm)}^{(n)}\rho_{(\pm)}^{(n)5}
       & \sqrt{2}\frac{1}{\sqrt{2}} f_{(\mp,\pm)}^{(n)}\rho_{(\pm)}^{(n)6} 
        & \pm\sqrt{2}\eta_n f_{(\pm,\pm)}^{(n)}D_{56(\pm)}^{\ast(n)3} \\ \hline
  \pm\sqrt{2}\eta_n f_{(\pm,\pm)}^{(n)}D_{56(\pm)}^{\ast(n)1}  
     & \pm\sqrt{2}\eta_n f_{(\pm,\pm)}^{(n)}D_{56(\pm)}^{\ast(n)2} 
       & \pm\sqrt{2}\eta_n f_{(\pm,\pm)}^{(n)}D_{56(\pm)}^{\ast(n)3}   
         & \sqrt{6}\frac{1}{\sqrt{2}} f_{(\mp,\pm)}^{(n)}\chi_{(\pm)}^{(n)} \\
  \ea \right). \no
 \eea

The diagonalized mass terms are 
 \bea
 \lag_4 &\supset&
  -\sum_{n=-\infty}^{\infty}\left[
            \sum_{i=1}^{3}\frac{n+\alpha}{R} \overline{\Psi}^{(n)i}_{(\pm)}\Psi^{(n)i}_{(\pm)}
           + \sum_{i=4}^{7}\frac{n+2\alpha}{R} \overline{\Psi}^{(n)i}_{(\pm)}\Psi^{(n)i}_{(\pm)}
            +\sum_{i=7}^{14}\frac{n+\alpha+1/2}{R} \overline{\Psi}^{(n)i}_{(\pm)}\Psi^{(n)i}_{(\pm)}            
            \right.\hspace{200pt}\non 
    &&        \left.
            +\frac{n+2\alpha+1/2}{R} \overline{\Psi}^{(n)15}_{(\pm)}\Psi^{(n)15}_{(\pm)}
            +\frac{n+3\alpha+1/2}{R} \overline{\Psi}^{(n)16}_{(\pm)}\Psi^{(n)16}_{(\pm)}\right] \non
 && -\sum_{n=0}^{\infty}\left[\sum_{i=17}^{32}\frac{n}{R} \overline{\Psi}^{(n)i}_{(\pm)}\Psi^{i(n)}_{(\pm)}
      +\sum_{i=33}^{40}\frac{n+1/2}{R} \overline{\Psi}^{(n)i}_{(\pm)}\Psi^{(n)i}_{(\pm)}\right]\non
 \eea
and the corresponding mass eigenstates are given by
 \bea
 &&\Psi^{(n)\{1,2,3\}}_{(\pm)}=\eta_n\left(\omega^{(n)\{2,4,6\}}_{(\pm)}
                                                                -Q^{\ast(n)\{1,3,5\}}_{56(\pm)}\right), \non
 &&  \Psi^{(n)\{4,5,6\}}_{(\pm)}=
                     \eta_n\left(Q^{\ast(n)\{2,4,6\}}_{56(\pm)}
                          -\frac{1}{\sqrt{2}}\left(\omega^{(n)\{7,8,9\}}_{(\pm)}
                                                     -D^{\ast(n)\{1,2,3\}}_{56(\pm)}\right)\right), \non
 && \Psi^{(n)\{7-12\}}_{(\pm)}=\frac{1}{\sqrt{2}}\left(\theta^{(n)\{7-12\}}_{(\pm)}
                                                                     -\rho^{(n)\{1-6\}}_{(\pm)}\right), \quad 
   \Psi^{(n)13}_{(\pm)}=
          \frac{1}{\sqrt{2}}\left(\phi^{(n)2}_{(\pm)}
                                                        -\tau^{(n)1}_{(\pm)}\right), \non
 &&\Psi^{(n)14}_{(\pm)}=\frac{1}{2\sqrt{2}}\left(\sqrt{3}\left(\phi^{(n)4}_{(\pm)}
                                                                     -\chi^{(n)}_{(\pm)}\right)
                                                            -\left(\tau^{(n)3}_{(\pm)}
                                                                     -\nu^{(n)2}_{(\pm)}\right)\right), \non
&&   \Psi^{(n)15}_{(\pm)}=
                     \frac{1}{\sqrt{2}}\left(\tau^{(n)2}
                          -\frac{1}{\sqrt{2}}\left(\phi^{(n)3}_{(\pm)}
                                                        -\nu^{(n)1}_{(\pm)}\right)\right), \non
 &&\Psi^{(n)16}_{(\pm)}=\frac{1}{2\sqrt{2}}\left(\left(\phi^{(n)4}_{(\pm)}
                                                                     -\chi^{(n)}_{(\pm)}\right)
                                                            -\sqrt{3}\left(\tau^{(n)3}_{(\pm)}
                                                                     -\nu^{(n)2}_{(\pm)}\right)\right), \non 
 &&   \Psi^{(n)\{17,18,19\}}_{(\pm)}=\eta_n\left(\omega^{(n)\{7,8,9\}}_{(\pm)}
                                                                  +D^{\ast(n)\{1,2,3\}}_{56(\pm)}\right), \quad
 \Psi^{(n)\{20,21,22\}}_{(\pm)}=\omega^{(n)\{1,3,5\}}_{(\pm)}, \non 
&&   \Psi^{(n)23-32}_{(\pm)}=\zeta^{(n)1-10}_{(\pm)}, \quad
 \Psi^{(n)33}_{(\pm)}=\frac{1}{\sqrt{2}}\left(\phi^{(n)3}_{(\pm)}+\nu^{(n)1}_{(\pm)}\right), \quad 
   \Psi^{(n)\{34-39\}}_{(\pm)}=\theta^{(n)\{1-6\}}_{(\pm)}, \non
 && \Psi^{(n)40}_{(\pm)}=\phi^{(n)1}_{(\pm)}. 
 \eea



\begin{thebibliography}{100}

\bibitem{GH} 
  N.~S.~Manton,
  Nucl.\ Phys.\ B {\bf 158}, 141 (1979);
  D.~B.~Fairlie,
  Phys.\ Lett.\ B {\bf 82}, 97 (1979), 
  J.\ Phys.\ G {\bf 5}, L55 (1979);
  Y.~Hosotani,
  Phys.\ Lett.\ B {\bf 126}, 309 (1983), 
  Phys.\ Lett.\ B {\bf 129}, 193 (1983), 
  Annals Phys.\  {\bf 190}, 233 (1989).

\bibitem{1loop}
  I.~Antoniadis, K.~Benakli and M.~Quiros,
  New J.\ Phys.\  {\bf 3}, 20 (2001); 
  G.~von Gersdorff, N.~Irges and M.~Quiros,
  Nucl.\ Phys.\ B {\bf 635}, 127 (2002); 
 R.~Contino, Y.~Nomura and A.~Pomarol,
  Nucl.\ Phys.\ B {\bf 671}, 148 (2003); 
  C.~S.~Lim, N.~Maru and K.~Hasegawa,
    J.\ Phys.\ Soc.\ Jap.\  {\bf 77}, 074101 (2008).  
  
    
\bibitem{2loop}
  N.~Maru and T.~Yamashita,
  Nucl.\ Phys.\ B {\bf 754}, 127 (2006); 
  Y.~Hosotani, N.~Maru, K.~Takenaga and T.~Yamashita,
  Prog.\ Theor.\ Phys.\  {\bf 118}, 1053 (2007). 
 
 \bibitem{Higgsphys}
  C.~S.~Lim, N.~Maru and T.~Miura,
  PTEP {\bf 2015}, no. 4, 043B02 (2015); 
  K.~Hasegawa, C.~S.~Lim and N.~Maru,
  J.\ Phys.\ Soc.\ Jap.\  {\bf 85}, no. 7, 074101 (2016); 
Y.~Adachi and N.~Maru,
  Phys.\ Rev.\ D {\bf 98}, no. 1, 015022 (2018); 
  arXiv:1809.02748 [hep-ph].
 
\bibitem{GHUST} 
  C.~S.~Lim and N.~Maru,
  Phys.\ Rev.\ D {\bf 75}, 115011 (2007). 
  
    
\bibitem{diphoton}
  N.~Maru and N.~Okada,
  Phys.\ Rev.\  D {\bf 77}, 055010 (2008); 
  Phys.\ Rev.\ D {\bf 87}, no. 9, 095019 (2013); 
  arXiv:1310.3348 [hep-ph].
  
\bibitem{Maru}
  N.~Maru,
  Mod.\ Phys.\ Lett.\  A {\bf 23}, 2737 (2008).

  
  
\bibitem{GHUflavor}
  Y.~Adachi, C.~S.~Lim and N.~Maru,
  Phys.\ Rev.\  D {\bf 76}, 075009 (2007); 
    Phys.\ Rev.\  D {\bf 79}, 075018 (2009); 
  Nucl.\ Phys.\ B {\bf 839}, 52 (2010); 
  Phys.\ Rev.\  D {\bf 80}, 055025 (2009). 


\bibitem{GHUmixing} 
  Y.~Adachi, N.~Kurahashi, C.~S.~Lim and N.~Maru,
  JHEP {\bf 1011}, 150 (2010); 
  JHEP {\bf 1201}, 047 (2012); 
  Y.~Adachi, N.~Kurahashi, N.~Maru and K.~Tanabe,
  Phys.\ Rev.\ D {\bf 85}, 096001 (2012); 
  Y.~Adachi, N.~Kurahashi and N.~Maru,
  arXiv:1404.4281 [hep-ph].

\bibitem{triple} 
  Y.~Adachi and N.~Maru,
  PTEP {\bf 2016}, no. 7, 073B06 (2016). 
 
 \bibitem{Yukawa} 
 Y.~Adachi and N.~Maru,
  Eur.\ Phys.\ J.\ Plus {\bf 130}, no. 8, 168 (2015);
  arXiv:1501.06229 [hep-ph].
%
N.~Maru and N.~Okada,
  arXiv:1604.01150 [hep-ph];
 A.~Das, N.~Maru and N.~Okada,
  arXiv:1704.01353 [hep-ph].
 
\bibitem{GHDM}
M.~Regis, M.~Serone and P.~Ullio,
  JHEP {\bf 0703}, 084 (2007); 
%
G.~Panico, E.~Ponton, J.~Santiago and M.~Serone,
  Phys.\ Rev.\ D {\bf 77}, 115012 (2008); 
%
M.~Carena, A.~D.~Medina, N.~R.~Shah and C.~E.~M.~Wagner,
  Phys.\ Rev.\ D {\bf 79}, 096010 (2009);
%
Y.~Hosotani, P.~Ko and M.~Tanaka,
  Phys.\ Lett.\ B {\bf 680}, 179 (2009); 
%
  N.~Haba, S.~Matsumoto, N.~Okada and T.~Yamashita,
  JHEP {\bf 1003}, 064 (2010); 
%
  S.~Funatsu, H.~Hatanaka, Y.~Hosotani, Y.~Orikasa and T.~Shimotani,
  PTEP {\bf 2014}, 113B01 (2014);
%
N.~Maru, T.~Miyaji, N.~Okada and S.~Okada,
  JHEP {\bf 1707}, 048 (2017); 
%
 N.~Maru, N.~Okada and S.~Okada,
  Phys.\ Rev.\ D {\bf 96}, 115023 (2017); 
  Phys.\ Rev.\ D {\bf 98}, no. 7, 075021 (2018). 
  
  
\bibitem{LM} 
  C.~S.~Lim and N.~Maru,
  Phys.\ Lett.\ B {\bf 653}, 320 (2007). 

\bibitem{otherGGHU}
  G.~Burdman and Y.~Nomura,
  Nucl.\ Phys.\ B {\bf 656}, 3 (2003);
  N.~Haba, Y.~Hosotani, Y.~Kawamura and T.~Yamashita,
  Phys.\ Rev.\ D {\bf 70}, 015010 (2004); 
%
K.~Kojima, K.~Takenaga and T.~Yamashita,
  Phys.\ Rev.\ D {\bf 84}, 051701 (2011); 
  Phys.\ Rev.\ D {\bf 95}, no. 1, 015021 (2017); 
  JHEP {\bf 1706}, 018 (2017); 
  Y.~Hosotani and N.~Yamatsu,
  PTEP {\bf 2015}, 111B01 (2015); 
%
   A.~Furui, Y.~Hosotani and N.~Yamatsu,
  PTEP {\bf 2016}, no. 9, 093B01 (2016). 

\bibitem{CGM} 
  C.~Csaki, C.~Grojean and H.~Murayama,
  Phys.\ Rev.\ D {\bf 67}, 085012 (2003). 
   
\bibitem{SSS} 
  C.~A.~Scrucca, M.~Serone and L.~Silvestrini,
  Nucl.\ Phys.\ B {\bf 669}, 128 (2003). 
  
  
\bibitem{Kawamura} 
  Y.~Kawamura,
  Prog.\ Theor.\ Phys.\  {\bf 105}, 999 (2001). 



\bibitem{CCP} 
  G.~Cacciapaglia, C.~Csaki and S.~C.~Park,
  JHEP {\bf 0603}, 099 (2006). 
   
\bibitem{DDG} 
  K.~R.~Dienes, E.~Dudas and T.~Gherghetta,
  Phys.\ Lett.\ B {\bf 436}, 55 (1998); 
  Nucl.\ Phys.\ B {\bf 537}, 47 (1999). 

\bibitem{PDUED} 
  T.~Appelquist, B.~A.~Dobrescu, E.~Ponton and H.~U.~Yee,
  Phys.\ Rev.\ Lett.\  {\bf 87}, 181802 (2001).


\end{thebibliography}
\end{document}